\documentclass[12pt]{article}

\usepackage{amssymb}
\usepackage{amstext}

\def\be{\begin{equation}}

\def\ee{\end{equation}}

\def\ba{\begin{eqnarray}}

\def\ea{\end{eqnarray}}

\def \aa {\alpha}
\def \bb {\beta}
\def \gg {\gamma}
\def \dd {\delta}

\def \mm {\mu}
\def \nn {\nu}

\def \rr {\rho}

\def \lll {\Lambda}

\def \mul {\odot}
\def \dirac {\partial {\hspace*{-.11in} {\not}}\;\;}

\def \ba {\bar}

\def \2 {{1 \over 2}}
\def \3 {{1 \over 3}}
\def \4 {{1 \over 4}}
\def \5 {{1 \over 5}}
\def \6 {{1 \over 6}}
\def \7 {{1 \over 7}}
\def \8 {{1 \over 8}}
\def \9 {{1 \over 9}}
\def \0 { \infty}

\def \pa {\partial}

\def \qq {\qquad}

\begin{document}
\pagestyle{empty}
\begin{flushright}
\begin{tabular}{ll}
SWAT-03/368 & \\
QMUL-PH-03-02 & \\
hep-th/0303036 & \\
05/03/03 & \\ [.3in]
\end{tabular}
\end{flushright}
\begin{center}
{\Large {\bf{Geometric Second Order Field Equations
for General  Tensor Gauge Fields}}} \\ [.7in] {\large{P. de
Medeiros$\, {}^{1}$ and C. Hull$\, {}^{2}$}} \\ [.3in]
$\,{}^{1}$\, {\emph{Physics Department, University of Wales Swansea, \\
Singleton Park, Swansea SA2 8PP, U.K.}} \\ [.2in]
$\,{}^{2}$\, {\emph{Physics Department, Queen Mary, University of
London, \\ Mile End Road, London E1 4NS, U.K.}} \\ [.3in]
{\tt{p.de.medeiros@swan.ac.uk}} , {\tt{c.m.hull@qmul.ac.uk}} \\ [.5in]
{\large{\bf{Abstract}}} \\ [.2in]
\end{center}
Higher spin tensor gauge fields have natural gauge-invariant field equations
written in terms of generalised curvatures, but these are typically of
higher than second order in derivatives.
We construct geometric second order field equations and actions for
general higher spin boson fields, and  first order ones for
fermions, which are non-local but
which become local on gauge-fixing, or on introducing auxiliary fields.
This generalises the results of Francia and  Sagnotti to all
representations of the Lorentz group.
\clearpage
\pagestyle{plain}
\pagenumbering{arabic}


\section{Introduction}

Fields in higher spin representations of the Lorentz group arise in
a variety of contexts. In perturbative string theory
they are present in the spectrum of massive modes, and limits in which an infinite number of these become massless are of considerable interest, as they could lead to symmetric phases of string theory with an infinite-dimensional unbroken symmetry group. Interacting theories 
with infinite numbers of massless  higher spin fields with anti-de Sitter vacua have
been constructed in {\cite{keyVasiliev}}, {\cite{keyvaseq}}, but so far these 
have only been constructed in anti-de Sitter spaces of   dimensions $D\le 5$, and in
certain generalised spacetimes.
These are associated with higher spin algebras, and higher
spin superalgebras 
  have recently been constructed in dimensions $D \leq 7$ {\cite{keysezsun}}. 
The free covariant field theories for higher spin gauge fields
have been discussed in {\cite{keylabmor}}, {\cite{Labastida:1987kw}}, {\cite{keysieg}}, {\cite{keytsul}}. Tensor gauge fields in unusual representations of the
Lorentz group are also an inevitable consequence of dualising certain
conventional gauge theories. This was analysed in {\cite{keyHul}},
where the dual forms of linearised gravity were found in arbitrary
dimensions and duality was discussed for general tensor fields.

Recently, there has been considerable progress in the understanding of
free higher spin massless gauge fields. In {\cite{keythesis}},
{\cite{keydeMHul}}, covariant local free field equations and
gauge-invariant actions were given for gauge fields in $D$ dimensions
transforming in any representation of $GL(D,{\mathbb{R}})$. The field
equations were given in terms of field strengths so that they were
manifestly invariant under higher spin gauge transformations. However,
these field equations in general involved more than two derivatives of
the gauge field. (The field equations were second order only for those
gauge fields in representations corresponding to Young tableaux with no
more than two columns.) In {\cite{keyFraSag}}, second order field
equations were considered for gauge fields in completely symmetric
tensor representations.
On fixing some of the symmetries, these reduced to the local second
order field equations of Fronsdal {\cite{keyFro}}. However, the covariant form
of the field equations of {\cite{keyFraSag}} were written in terms of
invariant field strengths and were non-local, involving inverse
D' Alembertian operators. These non-localities can be removed, however,
by gauge-fixing, or by introducing auxiliary compensator
fields. This gauge-fixing has been described in detail in {\cite{keyFraSag2}} for totally symmetric tensors, where it was found that they indeed describe the same number of on-shell degrees of freedom as those derived using the Fronsdal formalism. The purpose of this note is to write down geometric second
order field equations for all representations by combining the two
approaches. The idea is simple; consider one of the geometric higher
derivative field equations in $D$ dimensions of {\cite{keythesis}},
{\cite{keydeMHul}}, which is of order $2n$ in derivatives. The free
field equation is the vanishing of a higher spin generalisation of the
Einstein tensor, $E =0$, where $E$ involves $2n$ partial derivatives of
the gauge field $A$. In a physical gauge (i.e. a transverse
traceless gauge) this reduces to ${\square}^n A=0$ where $\square :=
\partial_\mu \partial^\mu$. Then a suitable second order field equation
generalising that of {\cite{keyFraSag}} is ${1 \over {\square}^{n-1}} E=0$,
and in physical gauge this reduces to $\square  A=0$, as it should. This means
that the apparent non-locality of the equation can be eliminated by a suitable
gauge choice, as in {\cite{keyFraSag}}, and it can also be eliminated by
introducing auxiliary fields {\cite{keyFraSag}}. An analogous procedure, again
based on {\cite{keyFraSag}}, is presented for obtaining first
order field equations for spinor-valued fermionic fields in any
representation of the Lorentz group.

Consider for example  a fourth rank totally symmetric tensor gauge field
$A_{\mm\nn\rr\sigma}= A_{(\mm\nn\rr\sigma )}$ represented by  
the Young tableau
\, {\tiny \begin{tabular}{|c|c|c|c|}\hline
& & & \\ \hline
\end{tabular}} \,
which has the  gauge
transformation
%
%
\be
\dd A_{\mm\nn\rr\sigma} \;\; =\;\; 4\, {\partial}_{(\mm} \lll_{\nn\rr\sigma )}
\label{eq:1}
\ee
with totally symmetric tensor gauge parameter $\lll_{\nn\rr\sigma}$. The
gauge-invariant field strength is
%
%
\be
F_{\mm _1 \mm_2\nn_1\nn_2\rr_1\rr_2 \sigma_1 \sigma_2} \;\; =\;\;
\pa_{\mm_1}\pa_{\nn_1}\pa_{\rr_1}\pa_{\sigma_1} A_{\mm_2\nn_2\rr_2 \sigma_2} \;
-\; (\mm _1 \leftrightarrow \mm_2) \; -\; ...
\label{eq:2}
\ee
which is antisymmetrised on each index pair, so that
%
%
\be
F_{\mm _1 \mm_2\nn_1\nn_2\rr_1\rr_2 \sigma_1 \sigma_2} \;\; =\;\; F_{[\mm _1
\mm_2][\nn_1\nn_2][\rr_1\rr_2][ \sigma_1 \sigma_2]}
\label{eq:3}
\ee
This is the natural generalisation of the linearised Riemann tensor and is represented by the Young tableau
\, {\tiny \begin{tabular}{|c|c|c|c|}\hline
& & & \\ \hline
& & & \\ \hline
\end{tabular}} \,
. If
the free field equation is to come from varying a gauge-invariant
action with respect to $A_{\mm\nn\rr\sigma}$, it should be of the form
${\cal{G}}_{\mm\nn\rr\sigma}=0$ for some gauge-invariant, totally
symmetric tensor ${\cal{G}}_{\mm\nn\rr\sigma}$, which is the
generalisation of the Ricci tensor. The natural choice is to define
%
%
\be
G_{\mm_1\nn_1\rr_1 \sigma_1} \;\; :=\;\; \eta ^{\mm_2\nn_2}\eta
^{\rr_2 \sigma_2}F_{\mm _1\mm_2\nn_1\nn_2\rr_1\rr_2 \sigma_1 \sigma_2}
\label{eq:4}
\ee
Then $G_{( \mm_1\nn_1\rr_1 \sigma_1 )}=0$ is a covariant local field
equation, but is fourth order in derivatives.
\footnote{The field equation $G_{\mm_1\nn_1\rr_1 \sigma_1}=0$ (without
symmetrisation) is also gauge-invariant but cannot be derived from a
gauge-invariant action without introducing extra fields.}
Here ${\eta}_{\mm\nn}$ is the background $SO(D-1,1)$-invariant
Minkowski metric, which is used to raise and lower indices. The second
order field equation of {\cite{keyFraSag}} is then given by
%
%
\be
{\cal{G}}_{\mm\nn\rr\sigma} \;\; :=\;\; {1 \over \square} G_{(
\mm\nn\rr\sigma )} \;\; =\;\; 0
\label{eq:5}
\ee
In the tranverse traceless gauge (which we will refer to as the \lq physical
gauge')
 %
%
\be
\pa ^\mm A_{\mm\nn\rr\sigma} \;\; =\;\; 0,\qq
\eta^{\mm\nn}A_{\mm\nn\rr\sigma} \;\; =\;\; 0
\label{eq:6}
\ee
the equation ({\ref{eq:5}}) reduces to $\square A_{\mm\nn\rr\sigma}=0$, as
required (see Appendix A for a discusssion of going to transverse traceless
gauge).

The generalised Einstein tensor is
%
%
\be
E_{\mm\nn\rr\sigma} \;\; :=\;\; G_{( \mm\nn\rr\sigma )} - {\eta}_{(\mm\nn} \,
G_{\rr\sigma )\aa\bb} \; {\eta}^{\aa\bb} + {1 \over 4}\, {\eta}_{(\mm\nn}
{\eta}_{\rr\sigma )} \, G_{\aa\bb\gg\dd} \; {\eta}^{\aa\bb} {\eta}^{\gg\dd}
\label{eq:7}
\ee
which satisfies the conservation equation
%
%
\be
{\partial}^{\mm} E_{\mm\nn\rr\sigma} \;\; \equiv\;\; 0
\label{eq:8}
\ee
identically. Then the action
%
%
\be
{\cal{S}}^{[1,1,1,1]}_{(1)} \;\; =\;\; -{1\over 24} \int d^D x \;
A^{\mm\nn\rr\sigma} {1\over \square} E_{\mm\nn\rr\sigma}
\label{eq:9}
\ee
is invariant under the gauge transformation ({\ref{eq:1}}) and its
variation gives the field equation ${1 \over \square} E_{\mu\nu\rho\sigma} =0$, which is equivalent to ({\ref{eq:5}}). This is a simple
modification of the action ${\cal{S}}^{[1,1,1,1]}$ for a type
$[1,1,1,1]$ tensor gauge field given in {\cite{keythesis}},
{\cite{keydeMHul}}, obtained by inserting the non-local operator
${\square}^{-1}$.

The generalisation to totally symmetric tensors $A_{\mm_1 ... \mm_s}$
of spin-$s$ is straightforward {\cite{keyFraSag}}. There is a field
strength
%
%
\be
F_{{\mm}^{1}_1 \, {\mm}^{1}_2 \, {\mm}^{2}_1 \, {\mm}^{2}_2 \, ...\,
{\mm}^{s}_1 \, {\mm}^{s}_2} \;\; =\;\; \pa_{{\mm}^{1}_1}
\pa_{{\mm}^{2}_1} ... \pa_{{\mm}^{s}_1} A_{{\mm}^{1}_2 \, {\mm}^{2}_2
\, ...\, {\mm}^{s}_2} \; -\; ( {\mm}^{1}_1 \leftrightarrow {\mm}^{1}_2
) \; -\; ...
\label{eq:10}
\ee
which is antisymmetrised on each of the $s$ index pairs so that
%
%
\be
F_{{\mm}^{1}_1 \, {\mm}^{1}_2 \, {\mm}^{2}_1 \, {\mm}^{2}_2 \, ...\,
{\mm}^{s}_1 \, {\mm}^{s}_2} \;\; = \;\; F_{[ {\mm}^{1}_1 \, {\mm}^{1}_2
]\, [ {\mm}^{2}_1 \, {\mm}^{2}_2 ]\, ... \, [ {\mm}^{s}_1 \,
{\mm}^{s}_2 ]}
\label{eq:11}
\ee
If $s$ is even, then contracting $F_{{\mm}^{1}_1 \, {\mm}^{1}_2 \,
{\mm}^{2}_1 \, {\mm}^{2}_2 \, ...\, {\mm}^{s}_1 \, {\mm}^{s}_2}$ over
$s/2$  pairs of indices with $s/2$ metric tensors $\eta
^{{\mm}^{i}_2 \, {\mm}^{j}_2}$ defines the gauge-invariant tensor
$G_{\mm_1 ... \mm_s}$. A generalised Ricci tensor $G_{( \mm_1 ... \mm_s
)}$ is then defined by total symmetrisation of all $s$ indices and
$G_{( \mm_1 ... \mm_s )} =0$ is a covariant field equation of order $s$
in derivatives. This is the $s$ derivative field equation of
{\cite{keythesis}}, {\cite{keydeMHul}} for an even spin-$s$ field. The
associated second order field equation of {\cite{keyFraSag}} is
%
%
\be
{\cal{G}}_{\mm_1 ... \mm_s} \;\; := \;\; {1 \over \square^{\, r}} G_{(
\mm_1 ... \mm_s )} \;\; =\;\; 0
\label{eq:12}
\ee
where $r= {s \over 2} -1$, and reduces to the Fronsdal equations
{\cite{keyFro}} on partial gauge fixing to traceless gauge
transformations {\cite{keyFraSag}}. These Fronsdal equations then
reduce further to $\square A_{\mm_1 ... \mm_s} =0$ on imposing   the
physical gauge conditions  $\partial^{\mm_1} A_{\mm_1 ... \mm_s} =0$ and
$\eta^{\mm_1 \mm_2} A_{\mm_1 \mm_2 ... \mm_s} =0$. This equation can be derived from a gauge-invariant non-local action of
the form
%
%
\be
{\cal{S}}^{[1,...,1]}_{(r)} \;\; =\;\; - {1 \over s!} \int d^D x\;
A^{\mm_1 ... \mm_s} {1 \over \square^{\, r}} E_{\mm_1 ... \mm_s}
\label{eq:13}
\ee
where $r= {s \over 2} -1$ and $E_{\mm_1 ... \mm_s}$ is the generalised
Einstein tensor given by shifting the traces of $G_{\mm_1 ... \mm_s}$
and Young symmetrising indices so that $\pa^{\mm_1} E_{\mm_1 ... \mm_s}
\equiv 0$ identically. The   construction of such \lq Einstein tensors' is discussed in   Appendix B.

For odd spins, we define a rank $s$ tensor
$G_{\mm_1 ... \mm_s}$ by contracting over ${(s+1) \over 2}$ pairs of
indices of $\partial_{\mm^{s+1}} F_{{\mm}^{1}_1 \, {\mm}^{1}_2 \,
{\mm}^{2}_1 \, {\mm}^{2}_2 \, ...\, {\mm}^{s}_1 \, {\mm}^{s}_2}$. Then
the field equation $G_{( \mm_1 ... \mm_s )} =0$ of {\cite{keythesis}},
{\cite{keydeMHul}} is of order $s+1$ in derivatives, and the second
order field equation of {\cite{keyFraSag}} is as in ({\ref{eq:12}}),
but with $r= {(s+1) \over 2} -1$. This equation follows from a
gauge-invariant action of the form ({\ref{eq:13}}), but with $r= {(s+1)
\over 2} -1$. The gauge-invariant tensor $E_{\mm_1 ... \mm_s}$ is again
defined by shifting the traces of $G_{\mm_1 ... \mm_s}$ and
symmetrising. In both even and odd spin cases,
the physical gauge field equation $\square \, A_{\mm_1 ... \mm_s}=0$ follows
from the     local action
%
%
\be
{\cal{S}}^{[1,...,1]}_{(r)} \;\; =\;\; - {1 \over s!} \int d^D x\;
A^{\mm_1 ... \mm_s} \square \, A_{\mm_1 ... \mm_s}
\label{eq:13a}
\ee

This generalises to gauge fields in arbitrary  representations of the
general linear group $GL(D,{\mathbb{R}})$. Consider for example the
Young tableau
\, {\tiny \begin{tabular}{|c|c|c|c|}\hline

& & & \\ \hline
   \\ \cline{1-1}
\end{tabular}} \,
corresponding to a tensor gauge field $A_{\mm_1 \mm_2 \nn\rr\sigma}$
satisfying
%
%
\be
A_{\mm_1 \mm_2\nn\rr\sigma} \; =\; A_{[\mm_1 \mm_2]\nn\rr\sigma}, \qq
A_{[\mm_1 \mm_2\nn]\rr\sigma} \; =\; 0 ,\qq A_{\mm_1 \mm_2\nn\rr\sigma} \;
=\; A_{\mm_1 \mm_2 (\nn\rr\sigma)}
\label{eq:14}
\ee
We refer to this Young tableau as type $[2,1,1,1]$, where the numbers
denote the length of the columns. The field strength is
%
%
\be
F_{\mm_1 \mm_2 \mm_3 \nn_1\nn_2\rr_1\rr_2 \sigma_1 \sigma_2} \;\; =\;\;
\pa_{\mm_1}\pa_{\nn_1}\pa_{\rr_1}\pa_{\sigma_1}
A_{\mm_2\mm_3\nn_2\rr_2 \sigma_2} \; -\; ( \nn _1 \leftrightarrow \nn_2) \;
-\; ...
\label{eq:15}
\ee
which is antisymmetrised on each of the four index types so that
%
%
\be
F_{\mm _1 \mm_2\mm_3\nn_1\nn_2\rr_1\rr_2 \sigma_1 \sigma_2} \;\; =\;\;
F_{[\mm_1\mm_2\mm_3][\nn_1\nn_2][\rr_1\rr_2][ \sigma_1 \sigma_2 ]}
\label{eq:16}
\ee
and is in a representation that corresponds to the Young tableau
\, {\tiny \begin{tabular}{|c|c|c|c|}\hline

& & & \\ \hline
& & & \\ \hline
   \\ \cline{1-1}
\end{tabular}} \,
of type $[3,2,2,2]$. Again, one can define a gauge-invariant tensor by
%
%
\be
G_{\mm_1\mm_2\nn_1\rr_1\sigma_1} \;\; := \;\; \eta^{\mm_3\nn_2}
\eta^{\rr_2\sigma_2} F_{\mm_1\mm_2\mm_3\nn_1\nn_2\rr_1\rr_2 \sigma_1 \sigma_2}
\label{eq:17}
\ee
This tensor is not irreducible under $GL(D,{\mathbb{R}})$ but an
associated Young tableau can be defined by the Young projection
${\cal{Y}}_{[2,1,1,1]}$ of ({\ref{eq:17}}) onto the irreducible
subspace of type $[2,1,1,1]$ tensors. This projected tensor corresponds
to the generalised Ricci tensor and  ${\cal{Y}}_{[2,1,1,1]} \circ
G_{\mm_1\mm_2\nn_1\rr_1\sigma_1} = 3\, \eta^{( \mm_3\nn_2}
\eta^{\rr_2\sigma_2 )} F_{\mm_1\mm_2\mm_3\nn_1\nn_2\rr_1\rr_2 \sigma_1 \sigma_2}
 =0$ is a covariant local fourth order
field equation, while  ${\cal{Y}}_{[2,1,1,1]} \circ \left( {1 \over
\square} G_{\mm_1\mm_2\nn_1\rr_1\sigma_1} \right) =0$ is the natural
candidate for a second order field equation. Again, an Einstein-type
tensor $E_{\mm_1\mm_2\nn_1\rr_1\sigma_1}$ of type $[2,1,1,1]$ can be
defined by shifting traces followed by Young projection of
$G_{\mm_1\mm_2\nn_1\rr_1\sigma_1}$ (see Appendix B), so that the second order field
equation above can be derived from the gauge-invariant action
%
%
\be
{\cal{S}}^{[2,1,1,1]}_{(1)} \;\; = \;\; -{1 \over 12} \int d^D x\;
A^{\mm_1\mm_2\nn_1\rr_1\sigma_1} {1 \over \square}
E_{\mm_1\mm_2\nn_1\rr_1\sigma_1}
\label{eq:17a}
\ee
The non-local field equations reduce to $\square \,
A_{\mm_1\mm_2\nn_1\rr_1 \sigma_1}=0$ 
in the physical gauge
%
%
\be
\partial^{\mm_1} A_{\mm_1\mm_2\nn_1\rr_1\sigma_1} \;\; =\;\; 0 \quad ,
\quad \partial^{\nn_1} A_{\mm_1\mm_2\nn_1\rr_1\sigma_1} \;\; =\;\; 0 \quad
, \quad \eta^{\mm_1 \nn_1} A_{\mm_1\mm_2\nn_1\rr_1\sigma_1} \;\; =\;\; 0
\label{eq:17c}
\ee
and this field equation follows from the
  local   action
%
%
\be
{\cal{S}}^{[2,1,1,1]}_{(1)} \;\; = \;\; -{1 \over 12} \int d^D x\;
A^{\mm_1\mm_2\nn_1\rr_1\sigma_1} \square \, A_{\mm_1\mm_2\nn_1\rr_1\sigma_1}
\label{eq:17b}
\ee

The procedure described above generalises to arbitrary tensor gauge
fields $A$ of type $[ p_1 ,..., p_N ]$ represented by a Young tableau
with $N$ columns, each of length $p_i$. The field strength $F$ is given
by acting with $N$ derivatives on $A$ and projecting onto the
representation $[ p_1 +1,..., p_N +1]$, so that in particular there is
an antisymmetrization on the indices in each column. A generalised
Ricci tensor of type $[p_1 ,..., p_N]$ is obtained by first taking
$N/2$ traces of $F$ if $N$ is even, or by taking an extra derivative and
taking $(N+1)/2$ traces of $\pa F$ if $N$ is odd. By summing over all
possible inequivalent ways of taking these traces one obtains the
generalised Ricci tensor whose vanishing defines the higher order field
equation in {\cite{keythesis}}, {\cite{keydeMHul}}. An associated
second order field equation is obtained by dividing the higher order
equation above by a suitable power $r$ of the D' Alembertian operator.
This power is $r=N/2-1$ for even $N$ or $r=(N+1)/2 -1$ for odd $N$. The
non-local gauge-invariant action from which these equations derive
takes the form $\int d^D x\; A \cdot {1 \over \square^{\, r}} E$ in
terms of the type $[ p_1 ,..., p_N ]$ generalised Einstein tensor $E$.
This non-local action is then replaced by
the local form $\int d^D x\; A
\cdot \square \, A$ to derive the field equation $\square A =0$ in an appropriate physical gauge.

Clearly, some care is needed in dealing with the index structure in
such exotic representations and in making the above prescription
precise. For $p$-form gauge fields, i.e. gauge fields of type $[p]$
with a tableau comprising a single column of length $p$, differential
forms provide the natural formalism to describe the theory. For $N>1$,
the generalisation to \lq multi-forms' provides the natural formalism
to describe the theory. In {\cite{keythesis}}, {\cite{keydeMHul}}, we
presented a theory of multi-forms and applied it to general gauge
theories, and in section 2 we review the parts of that which will be
used here. Earlier work on the application of differential analysis to
higher spin gauge theory appeared in {\cite{keyDVHen1}},
{\cite{keyDubVio}} and a similar multi-form construction was used
in {\cite{keyDVHen2}}, {\cite{keyBekBou}}.
While this paper was in preparation, 
{\cite{Bekaert:2003az}} appeared which gave further discussion of the
spin-three theory and briefly discussed field equations for general gauge
fields.


\section{Multi-form gauge theory}
We begin by reviewing the multi-form construction of gauge theories
with gauge potential transforming in an arbitrary irreducible
representation of $GL(D,{\mathbb{R}})$. Much of the material in this section is given in {\cite{keythesis}}, {\cite{keydeMHul}}. For a
discussion of Young tableaux, see {\cite{keyHam}}.


\subsection{Multi-forms}
A {\textit{multi-form}} of order $N$ is a tensor field $T$ that is an
element of the $N$-fold tensor product of $p_i$-forms (where
$i=1,...,N$), written
%
%
\be
X^{p_1 ,..., p_N} \;\; := \;\; \Lambda^{p_1} \otimes ... \otimes
\Lambda^{p_N}
\label{eq:4.1}
\ee
In general, this will be a reducible representation of
$GL(D,{\mathbb{R}})$. The components of $T$ are written
$T_{{\mu}^1_1 ...{\mu}^1_{p_1} ...{\mu}^i_1 ...{\mu}^i_{p_i} ...
{\mu}^N_{p_N} }$ and are taken to be totally antisymmetric in each set
of $\{ {\mu}^i \}$ indices, so that
%
%
\be
T_{{\mu}^1_1 ...{\mu}^1_{p_1} ...{\mu}^i_1 ...{\mu}^i_{p_i} ...
{\mu}^N_{1} ... {\mu}^N_{p_N} } \;\; =\;\; T_{[ {\mu}^1_1
...{\mu}^1_{p_1} ] ...[ {\mu}^i_1 ...{\mu}^i_{p_i} ]... [ {\mu}^N_{1}
... {\mu}^N_{p_N} ]}
\label{eq:4.2}
\ee
The generalisation of the operations on ordinary differential forms to
multi-forms of order $N$ over ${\mathbb{R}}^D$ is as follows.

The $\mul$-product is the natural generalisation of wedge product to multi-forms and is given by the map
%
%
\be
\mul : X^{p_1 ,...,p_N } \times X^{{p'}_1 ,...,{p'}_N } \rightarrow X^{p_1 +{p'}_1 ,...,p_N +{p'}_N }
\label{eq:4.3}
\ee
defined by the $N$-fold wedge product on the individual form subspaces.

There are $N$ inequivalent exterior derivatives
%

%
\be
d^{(i)} : X^{p_1 ,...,p_i ,...p_N } \rightarrow X^{p_1 ,...,p_i +1
,...,p_N }
\label{eq:4.4}
\ee
which are individually defined as the exterior derivatives acting on
the $\Lambda^{p_i}$ form subspaces. This definition implies
${d^{(i)}}^2 =0$ (with no sum over $i$) and that   $d^{(i)}$ commutes
with  $d^{(j)}$ for any $i,j$.
Since the multi-form space $X^{p_1 ,..., p_N }$ is isomorphic to the
multi-form space $X^{p_1 ,..., p_N ,0 }$ of one order higher, one
can introduce a further derivative operator on $X^{p_1 ,..., p_N }$, defined by
%
%
\be
\pa \; :=\; d^{(N+1)} : X^{p_1 ,..., p_N } \rightarrow X^{p_1 ,... ,p_N
,1}
\label{eq:4.4a}
\ee
This follows by taking the partial derivative of a multi-form $T \in
X^{p_1 ,...,p_N}$ to define an element in $X^{p_1 ,...,p_N,1}$ with
components $\pa _{\mm_1^{N+1}}T_{{\mu}^1_1 ...{\mu}^1_{p_1}
...{\mu}^i_1 ...{\mu}^i_{p_i} ... {\mu}^N_{p_N} }$.
One can also define the total derivative
%
%
\be
{\cal{D}} \; := \; \sum_{i=1}^{N} d^{(i)} \qquad , \qquad
   {\cal{D}}: X^{p_1 ,...,p_i ,...p_N } \rightarrow \sum_{i=1}^{N}
\oplus\, X^{p_1 ,...,p_i +1 ,...,p_N }
\label{eq:4.5}
\ee
which satisfies ${\cal{D}}^{N+1} =0$.

For representations of $SO(D-1,1) \subset GL(D,{\mathbb{R}})$ there are
$N$ inequivalent Hodge dual operations
%
%
\be
*^{(i)} : X^{p_1 ,...,p_i ,...p_N } \to X^{p_1 ,...,D- p_i ,...p_N }
\label{eq:4.7}
\ee
which are defined to act as the Hodge duals on the individual
$\Lambda^{p_i}$ form subspaces. This implies that ${*^{(i)}}^2 =
(-1)^{1+ p_i (D-p_i )}$ (with no sum over $i$) and that   $*^{(i)}$
commutes with  $*^{(j)}$ for any $i,j$.

This also allows one to define $N$ inequivalent \lq adjoint' exterior
derivatives
%
%
\be
{d^\dagger}^{(i)} \; :=\; {(-1)}^{1+D( p_i +1)} *^{(i)} d^{(i)} *^{(i)}
: X^{p_1 ,...,p_i ,...p_N } \rightarrow X^{p_1 ,...,p_i -1 ,...,p_N }
\label{eq:4.8}
\ee
This implies ${{d^\dagger}^{(i)}}^2 =0$ (with no sum over $i$) and any
two ${d^\dagger}^{(i)}$ commute. One can then define the Laplacian
operator
%
%
\be
\Delta \; :=\; d^{(i)} {d^\dagger}^{(i)} + {d^\dagger}^{(i)} d^{(i)} :
X^{p_1 ,...,p_i ,...p_N } \rightarrow X^{p_1 ,...,p_i ,...,p_N }
\label{eq:4.9}
\ee
with no sum over $i$. The action of the Laplacian
operator on multi-form $T \in X^{p_1 ,...,p_N}$ is independent of which
$i=1,...,N$ is chosen for $\Delta$. This can be seen in component form
since ${( \Delta T)}_{{\mu}^1_1 ...{\mu}^1_{p_1} ...{\mu}^i_1
...{\mu}^i_{p_i} ... {\mu}^N_{p_N} } = \square \, T_{{\mu}^1_1
...{\mu}^1_{p_1} ...{\mu}^i_1 ...{\mu}^i_{p_i} ... {\mu}^N_{p_N} }$,
where $\square := \partial_\mu \partial^\mu$ is the D' Alembertian
operator on ${\mathbb{R}}^{D-1,1}$.

There exist $N(N-1)/2$ inequivalent trace operations
%
%
\be
\tau^{(ij)} : X^{p_1 ,...,p_i ,..., p_j ,...p_N } \rightarrow X^{p_1
,...,p_i -1,..., p_j -1,...,p_N }
\label{eq:4.10}
\ee
defined as the single trace between the $\Lambda^{p_i}$ and
$\Lambda^{p_j}$ form subspaces using the Minkowski metric
$\eta^{{\mu}^i_1 {\mu}^j_1}$. This allows one to define two
inequivalent \lq dual-trace' operations
%
%
\be
\sigma^{(ij)} \; :=\; {(-1)}^{1+D( p_i +1)} *^{(i)} \tau^{(ij)} *^{(i)}
: X^{p_1 ,...,p_i ,..., p_j ,...p_N } \rightarrow X^{p_1 ,...,p_i
+1,..., p_j -1,...,p_N }
\label{eq:4.11}
\ee
and
%
%
\be
{\tilde{\sigma}^{(ij)}} \; :=\; {(-1)}^{1+D( p_j +1)} *^{(j)}
\tau^{(ij)} *^{(j)} : X^{p_1 ,...,p_i ,..., p_j ,...p_N } \rightarrow
X^{p_1 ,...,p_i -1,..., p_j +1,...,p_N }
\label{eq:4.12}
\ee
associated with a given $\tau^{(ij)}$ (with no sum over $i$ or $j$).
Notice that ${\tilde{\sigma}}^{(ij)} = \sigma^{(ji)}$ since
$\tau^{(ij)} = \tau^{(ji)}$. This implies that the components ${( \sigma^{(ij)} T)}_{{\mu}^1_1 ... [ {\mu}^i_{1} ...{\mu}^i_{p_i} {\mu}^j_{1} ]  ... [ {\mu}^j_2 ... {\mu}^j_{p_j} ] ... {\mu}^N_{p_N} }$ are equal to ${(-1)}^{p_i +1} ( p_i +1) \, T_{{\mu}^1_1 ...[ {\mu}^i_{p_1} ... {\mu}^i_{p_i} | ... | {\mu}^j_{1} ] {\mu}^j_{2} ... {\mu}^j_{p_j} ... {\mu}^N_{p_N} }$.

There are $N(N-1)/2$ inequivalent involutions
%
%
\be
t^{(ij)} : X^{p_1 ,...,p_i ,..., p_j ,...p_N } \rightarrow X^{p_1
,...,p_j ,..., p_i ,...,p_N }
\label{eq:4.13}
\ee
defined by exchange of the $\Lambda^{p_i}$ and $\Lambda^{p_j}$ form
subspaces in the tensor product space. The components ${( t^{(ij)} T )}_{{\mu}^1_1 ... {\mu}^j_1 ... {\mu}^j_{p_j} ... {\mu}^i_1 ... {\mu}^i_{p_i} ... {\mu}^N_{p_N} }$ are proportional to $T_{{\mu}^1_1 ...[ {\mu}^j_{1} ... {\mu}^j_{p_j} ] [ {\mu}^i_1 ... {\mu}^i_{p_i - p_j} | ... | {\mu}^i_{p_i - p_j +1} ... {\mu}^i_{p_i} ] ... {\mu}^N_{p_N} }$ (assuming $p_i \geq p_j$).

There are also $N(N-1)/2$ inequivalent product operations
%
%
\be
\eta^{(ij)} : X^{p_1 ,...,p_i ,..., p_j ,...p_N } \rightarrow X^{p_1 ,...,p_i +1,..., p_j +1,...,p_N }
\label{eq:4.14}
\ee
defined as the $\mul$-product with the $SO(D-1,1)$ metric $\eta$ (understood as an order $N$ multi-form with all columns of zero length except $p_i = p_j =1$, corresponding to a
$[1,1]$ bi-form
 in the $\Lambda^{p_i} \otimes \Lambda^{p_j}$ subspace), such that $\eta^{(ij)} T \equiv \eta \mul T$ for any $T \in X^{p_1 ,..., p_N}$. The components ${( \eta^{(ij)} T )}_{{\mu}^1_1 ... {\mu}^i_1 ... {\mu}^i_{p_i +1} ... {\mu}^j_1 ... {\mu}^j_{p_j +1} ... {\mu}^N_{p_N} }$ are equal to $( p_i +1) ( p_j +1) \, \eta_{\mu^i_1 \mu^j_1} T_{{\mu}^1_1 ... {\mu}^i_2 ... {\mu}^i_{p_i +1} ... {\mu}^j_2 ... {\mu}^j_{p_j +1} ... {\mu}^N_{p_N} }$ (with implicit antisymmetrisation on the $( p_i +1)$ $\mu^i$ and $( p_j +1)$ $\mu^j$ indices seperately).


\subsection{Irreducible representations and Young tableaux}

The space of multi-forms $X^{p_1 ,..., p_N}$ is in general a reducible
representation of $GL(D,{\mathbb{R}})$. Each irreducible representation
of $GL(D,{\mathbb{R}})$ is associated with a Young tableau. Consider
the representation associated with a Young tableau with $N$ columns and
with $p_i$ cells in the $i$th column (it is assumed $p_i \geq
p_{i+1}$); we denote this representation as $[ p_1 ,..., p_N ]$. A
tensor $A$ in this representation is a multi-form $A \in X^{p_1 ,...,
p_N}$ satisfying
%
%
\be
\sigma^{(ij)} A \;\; =\;\; 0
\label{eq:4.15}
\ee
for any $j > i$ and also satisfying $t^{(ij)} A =A$ if the $i$th and
$j$th columns are of equal length, $p_i = p_j$
{\cite{keyHam}}. The projector from $X^{p_1 ,...,p_N}$ onto this
irreducible tensor representation $X^{[p_1 ,...,p_N ]}$ is the
Young symmetriser ${\cal{Y}}_{[p_1 ,...,p_N ]}$.

For example, consider a rank $s$ multi-form in $X^{1,1,...,1}$ (with
$N=s$ and all $p_i =1$). This is the space of all rank $s$ tensors
$T_{\mm_1 \mm_2 ... \mm_s}$, with no index symmetry properties and so
is a reducible representation of $GL(D,{\mathbb{R}})$. The projector
${\cal{Y}}_{[s,0,...,0]}$ takes this to the space of totally
antisymmetric tensors
$T_{\mm_1 \mm_2 ...\mm_s} = T_{[ \mm_1 \mm_2 ...\mm_s ]}$, while
${\cal{Y}}_{[1,1,...,1]}$ takes this to the space of totally symmetric
tensors $T_{\mm_1 \mm_2 ... \mm_s} = T_{( \mm_1 \mm_2 ...\mm_s )}$. The
full set of irreducible representations are obtained by acting on
$X^{1,1,...,1}$ with all projectors ${\cal{Y}}_{[p_1 ,...,p_s ]}$ with
$p_i
\geq p_{i+1}$ satisfying $\sum_{i=1}^{s} p_i =s$.


\subsection{Multi-form gauge theory}
Consider a gauge potential $A$ that is a tensor in the $[ p_1 ,..., p_N
]$ irreducible representation of $GL(D,{\mathbb{R}})$ whose components
have the index symmetry of an $N$-column Young tableau with $p_i$
cells in the $i$th column (with $p_i \geq p_{i+1}$). The natural gauge
transformation for this object is given by {\cite{keydeMHul}}
%
%
\begin{equation}
\delta A \;\; = \;\; {\cal{Y}}_{[ p_1 ,..., p_N ]} \circ \left( \,
\sum^N_{i=1}\; d^{(i)} \alpha_{(i)}^{p_1 ,...,p_i -1 ,...,p_N } \,
\right)
\label{eq:4.16}
\end{equation}
for any gauge parameters $\alpha_{(i)}^{p_1 ,..., p_i -1,..., p_N } \in
X^{p_1 ,..., p_i -1,..., p_N }$.

The associated field strength $F$ is a type $[ p_1 +1 ,..., p_N +1]$
tensor given by
%
%
\begin{equation}
F \;\; = \;\; \left( \prod_{i=1}^{N} \; d^{(i)} \right) A \;\; =
\;\; {1 \over N!} \, {\cal{D}}^N A
\label{eq:4.17}
\end{equation}
which is invariant under ({\ref{eq:4.16}}). The first expression is
unambiguous since all $d^{(i)}$ commute.

From the generalised Poincar\'{e} lemma in {\cite{keyDVHen1}},
{\cite{keyDubVio}}, {\cite{keyDVHen2}} it follows that any type $[ p_1
+1,..., p_N +1 ]$ tensor $F$ satisfying $d^{(i)} F=0$ for all $i$ can
be written as in ({\ref{eq:4.17}}) for some type $[ p_1 ,..., p_N ]$
potential $A$. The field strength $F$ satisfies second Bianchi
identities
%
%
\begin{equation}
d^{(i)} F\;\; =\;\; 0
\label{eq:4.18}
\end{equation}
and the first Bianchi identities
%
%
\begin{equation}
\sigma^{(ij)} F\;\; =\;\; 0
\label{eq:4.19}
\end{equation}
for any $j > i$.

Considering the irreducible representations of $GL(D,{\mathbb{R}})$
above to be reducible representations of the $SO(D-1,1)$ Lorentz
subgroup allows the construction of a gauge-invariant action functional
from which physical equations of motion can be obtained.

For $N$ odd, the natural field equation for a general type $[ p_1 ,...,
p_N ]$ gauge potential $A$ is given by
%
%
\begin{equation}
\sum_{I \in S_N}
\tau^{( i_1 i_2 )} ... \tau^{( i_N \, N+1)} \pa  F \;\; =\;\; 0
\label{eq:4.24}
\end{equation}
where the sum is on the labels $I = ( i_1 ... i_N )$ whose values vary
over all permutations of the set $(1...N)$. The $(N+1)$th label is not
included in the sum. The fact that the Young projection ${\cal{Y}}_{[ p_1 ,..., p_N ,0]}$ onto the irreducible $[ p_1 ,..., p_N ,0]$ tensor
 subspace is not required in this expression is shown in Appendix B. For $N$ even, the field equation for
a type $[ p_1 ,..., p_N ]$ gauge potential $A$ is given by
%
%
\begin{equation}
\sum_{I \in S_N} \tau^{( i_1
i_2 )} ... \tau^{( i_{N-1} i_N )} F \;\; =\;\; 0
\label{eq:4.25}
\end{equation}
where the sum here is on all the labels $I = ( i_1 ... i_N )$ whose
values vary over all permutations of the set $(1...N)$. The Young
projection ${\cal{Y}}_{[ p_1 ,..., p_N ]}$ is again unnecessary.

If
these field equations can be derived from a gauge-invariant action, it must be of the form
%
%
\begin{equation}
{\cal{S}}^{[ p_1 ,..., p_N ]} \;\; =\;\; - \left( \prod_{i=1}^{N}
\frac{1}{p_i !} \right) \int d^D x \; A^{{\mu}^1_1 ...{\mu}^1_{p_1}
...{\mu}^i_1 ...{\mu}^i_{p_i} ...{\mu}^N_{p_N} } E_{{\mu}^1_1
...{\mu}^1_{p_1} ...{\mu}^i_1 ...{\mu}^i_{p_i} ...{\mu}^N_{p_N}}
\label{eq:4.26}
\end{equation}
in terms of the type $[ p_1 ,..., p_N ]$ gauge potential $A$ and some
gauge-invariant field equation tensor $E$ involving $N$ partial
derivatives on $A$ for even $N$ (or $N+1$ derivatives for odd $N$).
Gauge invariance of ({\ref{eq:4.26}}) requires that $E$
should satisfy the $N$ conservation conditions $\partial^{\mu^i_1}
E_{{\mu}^1_1 ...{\mu}^1_{p_1} ...{\mu}^i_1 ...{\mu}^i_{p_i}
...{\mu}^N_{p_N}} \equiv 0$ identically for $i=1,...,N$. For $N$ even, the leading term in $E$   involves $N/2$ traces of the
field strength $F$ of $A$ and is given by the sum over all permutations of $N$ labels of
the term $F_{ {\mu}^1_1 ... {\mu}^1_{p_1 +1} ...
{\mu}^N_{p_N +1} } \; \eta^{\mu^1_{1} \, \mu^2_{1}} ...
\eta^{\mu^{N-1}_{1} \, \mu^N_{1}}$. The correction terms then consist
of   further traces (appropriately symmetrised) with coefficients
fixed by overall conservation of $E$ {\cite{keydeMHul}},
so that the field equation $E=0$ is a linear combination of the field equation given above and its multiple trace parts.
 For $N$ odd one can consider the
potential to be a type $[ p_1 ,..., p_N ,0]$ tensor of even order $N+1$
whose field strength $\pa F$ is a type $[ p_1 +1,..., p_N +1 ,1]$
tensor. The construction of $E$ is then the same as for the even $N$
case. The explicit form of $E$ is discussed in Appendix B, where it is constructed explicitly 
for bi-forms and totally symmetric tensors, and a general form is conjectured.

For general $N$ the local  gauge-invariant field equations above are
$N$th order in derivatives for even $N$ and of order $N+1$ for odd $N$.
Consequently they are higher derivative equations of motion for higher
spin fields with $N>2$ involving more than two partial derivatives of
the gauge field. In the next section we give a method for obtaining
second order field equations for such higher spin  tensor fields with
arbitrary $N$. This construction was described in {\cite{keyFraSag}}
for the case of totally symmetric spin-$s$ gauge fields. In our
notation such totally symmetric spin-$s$ fields are tensors of type
$[1,...,1]$ with $N=s$ entries.


\section{Second order field equations}
The local gauge-invariant action ({\ref{eq:4.26}}) for a general type
$[ p_1 ,..., p_N ]$ gauge potential $A$ can be modified by the
insertion of a negative power of the D' Alembertian scalar operator
$\square$. The resulting non-local action is given by
%
%
\begin{equation}
{\cal{S}}_{(r)}^{[ p_1 ,..., p_N ]} \;\; =\;\; - \left( \prod_{i=1}^{N}
\frac{1}{p_i !}
\right) \int d^D x \; A^{{\mu}^1_1 ...{\mu}^1_{p_1} ...{\mu}^i_1
...{\mu}^i_{p_i}
...{\mu}^N_{p_N} } {1 \over {\square}^{\, r}} E_{{\mu}^1_1
...{\mu}^1_{p_1} ...{\mu}^i_1
...{\mu}^i_{p_i} ...{\mu}^N_{p_N}}
\label{eq:1.1}
\end{equation}
and is gauge-invariant for any power $r$.
Formally, this gives field equations of order $N-2r$ for $N$ even or
$N+1-2r$ for $N$ odd, so that choosing $r = {N
\over 2} -1$ for $N$ even and $r = {(N+1) \over 2} -1$ for $N$ odd
gives second order field equations. For any $r$, these field equations are covariant and gauge-invariant, but are
non-local in general. We will show that the second order field equations become local in physical gauge, and it is to be expected that the non-localities could instead be eliminated by the introduction of auxiliary fields, as shown for  the spin-three case   in  {\cite{keyFraSag}}.
For the case in which $r$ is chosen to make the field equation zero'th order, the field equations 
imply the fields are pure gauge.

 Choosing   $r = {N
\over 2} -1$ for $N$ even and $r = {(N+1) \over 2} -1$ for $N$ odd,
gives  gauge-invariant field equations derived from
({\ref{eq:1.1}})  that are
of second order and are given by the non-local expressions
%
%
\begin{eqnarray}
{\cal{G}}^{(0)} &:=& \sum_{I
\in S_N} \tau^{( i_1 i_2
)} ... \tau^{( i_{N-1} i_N )} {1 \over {\square}^{\, {N \over 2} -1}} F \;\; =\;\; 0 \label{eq:1.2} \\ [.1in]
{\cal{G}}^{(1)} &:=& \sum_{I \in S_N}
\tau^{( i_1 i_2 )} ... \tau^{( i_N \, N+1)} {1 \over {\square}^{\,
{(N+1) \over 2} -1}} \pa  F \;\; =\;\; 0 \label{eq:1.3}
\end{eqnarray}
for $N$ even and odd respectively. The Young projections ${\cal{Y}}_{[ p_1 ,..., p_N ]}$ and ${\cal{Y}}_{[ p_1 ,..., p_N ,0]}$ in ({\ref{eq:1.2}}) and ({\ref{eq:1.3}}) are not necessary, following the theorem in Appendix B. These equations correspond to those proposed in
{\cite{keyFraSag}} for the case of a spin-$s$ gauge field with $N=s$
and all $p_i =1$. The non-local action ({\ref{eq:1.1}}) is then replaced by the
local action
%
%
\begin{equation}
{\cal{S}}_{(r)}^{[ p_1 ,..., p_N ]} \;\; =\;\; - \left( \prod_{i=1}^{N}
\frac{1}{p_i !}
\right) \int d^D x \; A^{{\mu}^1_1 ...{\mu}^1_{p_1} ...{\mu}^i_1
...{\mu}^i_{p_i}
...{\mu}^N_{p_N} } \square \, A_{{\mu}^1_1 ...{\mu}^1_{p_1}
...{\mu}^i_1 ...{\mu}^i_{p_i}
...{\mu}^N_{p_N} }
\label{eq:1.1a}
\end{equation}
in the   physical gauge
%
%
\begin{equation}
{d^{\dagger}}^{(i)} A \;\; =\;\; 0 \quad\quad , \quad\quad \tau^{(ij)}
A \;\; =\;\; 0
\label{eq:1.1b}
\end{equation}
for any $i,j = 1,...,N$.

The second order, non-local field equations above are not unique.
Define $F^{(m)} := {\pa}^{m} F$ to be the order $N+m$ tensor associated
with the canonical field strength tensor $F$ of order $N$ obtained by
acting on $F$ with $m$ partial derivatives. Then write
%
%
\begin{eqnarray}
{\cal G}^{(2n)} &:=& \sum_{I \in S_{N+2n}} \tau^{( i_1 i_2 )} ... \tau^{( i_{N-1+2n} \;
i_{N+2n} )} {1 \over {\square}^{\, {N \over 2}
-1+n}} F^{(2n)}  \label{eq:1.4} \\ [.1in]
{\cal G}^{(2n+1)} &:=& \sum_{I \in S_{N+2n}} \tau^{( i_1 i_2 )} ... \tau^{( i_{N+2n} \,
N+1+2n)} {1 \over {\square}^{\, {(N+1) \over 2} -1+n}} F^{(2n+1)} \label{eq:1.5}
\end{eqnarray}
for the case of $N$ even and odd respectively. The Young projections ${\cal{Y}}_{[ p_1 ,..., p_N ,0,...,0]}$ (with $2n$ zeros) and ${\cal{Y}}_{[ p_1 ,..., p_N ,0,0,...,0]}$ (with $2n+1$ zeros) in ({\ref{eq:1.4}}) and ({\ref{eq:1.5}}) respectively, are not necessary. It is clear by construction that ({\ref{eq:1.4}}) and
({\ref{eq:1.5}}) are related to the original field equation tensors in
({\ref{eq:1.2}}) and ({\ref{eq:1.3}}) and the equation  ${\cal
G}^{(m)}=0$ also reduces to $\square A =0$ if one imposes  the physical gauge
conditions ({\ref{eq:1.1b}}). The field equations given by the vanishing of ({\ref{eq:1.4}}) and
({\ref{eq:1.5}}) are more restrictive than ({\ref{eq:1.2}}) and
({\ref{eq:1.3}}) in the sense that the gauge-invariant tensors in
({\ref{eq:1.4}}) and ({\ref{eq:1.5}}) vanish as a consequence of
({\ref{eq:1.2}}) and ({\ref{eq:1.3}}) though the converse statement is
not true
\footnote{A simple example to illustrate this fact is for linearised
gravity where ${\cal{G}}^{(0)}_{\mm\nn} = R_{\mm\nn}$ is the Ricci
tensor and ${\cal{G}}^{(2)}_{\mm\nn} = R_{\mm\nn} - 2 {1 \over \square}
\partial^{\rr} \partial^{\sigma} R_{\rr\mm\sigma\nn}$ where
$R_{\mm\nn\rr\sigma}$ is the full linearised Riemann tensor. It is clear
that the non-trivial Einstein equation $R_{\mm\nn} =0$ in $D \geq 4$
implies the secondary field equation $\partial^{\mm} R_{\mm\nn\rr\sigma}
=0$ (by tracing the second Bianchi identity $\partial_{[ \alpha}
R_{\mm\nn ] \rr\sigma} =0$) so that ${\cal{G}}^{(0)} =0$ implies
${\cal{G}}^{(2)} =0$ but not vice versa. This structure follows in the
general theory where one expands a given field equation of level $m$ (i.e. ${\cal{G}}^{(m)} =0$) and finds only lower field equations of levels $<m$ and their various
\lq secondary' field equations derived using these lower level
equations and the associated Bianchi identities.}
. One of these associated field equations is noted in
{\cite{keyFraSag}} for the case of a spin-3 field. In our framework
this example corresponds to the case in which $N=3$, $p_1 = p_2 = p_3
=1$ and $n=1$. As noted in {\cite{keyFraSag}}, this second order field
equation ${\cal G}^{(3)} =0$ is simply related to a linear combination
of the equation ${\cal G}^{(1)} =0$ and its trace.
In general, one can consider field equations given by linear
combinations of these tensors,
$\sum _n a_n \, {\cal G}^{(2n)} =0$ or $\sum _n a_n \, {\cal G}^{(2n+1)} =0$
for some coefficients $a_n$, but these will generally be more restrictive than ({\ref{eq:1.2}}) and ({\ref{eq:1.3}}).


\section{Connections}

The linearised
Riemann tensor can be written as the (appropriately symmetrised) single
derivative of a first order linearised connection, such that
$R_{\mu\nu\rho\sigma} = 4 \partial_{[ \mu} \Gamma_{\nu ][ \rho\sigma
]}$ where $\Gamma_{\nu\rho\sigma} = {1 \over 2} ( \partial_{\nu}
h_{\rho\sigma} + \partial_{\rho} h_{\nu\sigma} - \partial_{\sigma}
h_{\nu\rho} )$. More generally, {\cite{keyFraSag}}, {\cite{keydWFre}}, the field strength of a 
general spin-$s$ gauge
field can be written as a
derivative of a rank $2s-1$ linear connection involving $s-1$
derivatives of the spin-$s$ gauge field. We show that such a linear
connection structure arises for general tensor gauge theories.

The type $[ p_1 +1,..., p_N +1]$ tensor field
strength associated with an arbitrary type $[ p_1 ,..., p_N ]$ gauge
potential $A$ satisfies $d^{(i)} F =0$ for all $i$. For any given $i$,
this implies that $F$ is $d^{(i)}$-exact, so that
%
%
\begin{equation}
F \;\; = \;\; d^{(i)} {\tilde{\Gamma}}_{(i)}
\label{eq:2.1}
\end{equation}
(with no sum over $i$) where  ${\tilde{\Gamma}}_{(i)} \in X^{p_1
+1,..., p_i ,..., p_N +1}$ is defined by
%
%
\begin{equation}
{\tilde{\Gamma}}_{(i)} \;\; := \;\; \left( \prod_{j \neq i} \; d^{(j)}
\right) A
\label{eq:2.2}
\end{equation}
Notice that ${\tilde{\Gamma}}_{(i)}$ is a multi-form involving $N-1$
derivatives of $A$ but is not $GL(D,{\mathbb{R}})$-irreducible in 
general. Under
the gauge transformation ({\ref{eq:4.16}}),
${\tilde{\Gamma}}_{(i)}$ transforms as
%
%
\begin{equation}
\delta {\tilde{\Gamma}}_{(i)} \;\; = \;\; \left( \prod_{j=1}^{N} \;
d^{(j)} \right) \alpha_{(i)}^{p_1 ,..., p_i -1,..., p_N }
\label{eq:2.3}
\end{equation}
and is invariant under the transformations with parameter $\alpha_{(j)}$ for any $j \ne i$,
so that $F$ is gauge-invariant. All $N$ ${\tilde{\Gamma}}_{(i)}$ are
inequivalent if no two column lengths $p_i$ are equal. For any two columns
$i$ and $j$ of equal length (with $p_i = p_j$) then
${\tilde{\Gamma}}_{(i)}$ and ${\tilde{\Gamma}}_{(j)}$ are equivalent
under transposition, in the sense that ${\tilde{\Gamma}}_{(i)} =
t^{(ij)} {\tilde{\Gamma}}_{(j)}$ (with no sum over $j$). This explains
why only a single linear connection is realised in {\cite{keyFraSag}},
{\cite{keydWFre}} for a spin-$s$ gauge field with all $p_i =1$.

Note that for gravity, the linearised Christoffel connection
$\Gamma$ is related to the bi-form connection ${\tilde{\Gamma}}$ defined in 
this way by $2 \Gamma_{\mu [ \nu\rho ]} = -2 h_{\mu [ \nu , \rho ]} \equiv {\tilde{\Gamma}}_{\mu\nu\rho} \in X^{1,2}$
and is distinguished by its transformation property
$\delta \Gamma_{\mu\nu\rho} = \partial_{\mu}
\partial_{\nu} \xi_{\rho}$ under $\delta h_{\mu\nu} = 2 \partial_{( \mu} \xi_{\nu )}$.
For totally symmetric tensors, generalisations of the linearised 
Christoffel connection $\Gamma$ were proposed in {\cite{keydWFre}}, and these are related to linear 
combinations of the ${\tilde{\Gamma}}$, but for the general case, the multi-forms ${\tilde{\Gamma}}_{(i)}$ seem to be more natural in the linear theory.

For a totally symmetric spin-$s$ field, there is in fact a hierarchy 
of connections
{\cite{keydWFre}}, and the same is true for general  $[ p_1 ,..., p_N ]$ tensor gauge fields $A$
with gauge-invariant field strength
$F = d^{(1)} ... d^{(N)} A$.
In addition to the $N$
multi-forms ${\tilde{\Gamma}}_{(i)}$ of order $N-1$ in derivatives
defined above, one can  define $N!/(N-k)!k!$ multi-forms
of order $N-k$ in derivatives as
%
%
\begin{equation}
{\tilde{\Gamma}}_{( i_1 ... i_k )} \;\; := \;\; \left( \prod_{i \not\in
( i_1 ,..., i_k )} \; d^{(i)} \right) A \;\; \in \;\; X^{p_1 +1,...,
p_{i_1} ,..., p_{i_k} ,..., p_N +1}
\label{eq:2.6}
\end{equation}
by   pulling off $k$ different exterior derivatives from the
definition of $F$ in all possible inequivalent ways. Consequently, at
the top of the hierarchy there is one ${\tilde{\Gamma}} = F$ and at the
bottom there is also one inequivalent ${\tilde{\Gamma}}_{( i_1 ... i_N
)} = A$. It is always possible to write $F$ in terms of any one of
these multiforms since
%
%
\begin{equation}
F \;\; =\;\; d^{({i_{1}})} ... d^{({i_k})} {\tilde{\Gamma}}_{( i_1 ...
i_k )}
\label{eq:2.7}
\end{equation}
Therefore, by construction, each multi-form ${\tilde{\Gamma}}_{( i_1
... i_k )}$ transforms in such a way that $F$ is invariant under gauge
transformation ({\ref{eq:4.16}}).


\section{First order fermionic field equations}

It is straightforward to generalise the analysis of a bosonic tensor 
gauge field
in the representation   $[ p_1 ,..., p_N ]$
to the case of a  fermionic spinor-valued tensor gauge field in the 
representation
${[ p_1 ,..., p_N ]}_S := [ p_1 ,..., p_N ] \otimes {\textsf{S}}$ of (the cover of) the Lorentz group, given
by the tensor product of the tensor representation with the Dirac spinor representation
$ {\textsf{S}}$ (with no constraints on traces or gamma-traces).

For a Dirac spinor $\psi \in X^{{[0]}_S}$,
the Dirac equation $\dirac\, \psi =0$ implies the
Klein-Gordon equation $\square\, \psi =0$ (where ${\dirac} := \gamma^{\mu} \partial_{\mu}$ and $\gamma^{\mu} \gamma^{\nu} + \gamma^{\nu} \gamma^{\mu} = 2 \eta^{\mu\nu}$) while the Dirac equation
can formally be obtained by acting on the Klein-Gordon equation 
$\square\, \psi =0$ with the
non-local operator ${1 \over \square} \dirac$.
Similarly for a Dirac spinor-valued vector (gravitino) $\psi \in X^{{[1]}_S}$, the
Rarita-Schwinger equation $\gamma^{\mu\nu\rho} \partial_{\nu}
\psi_{\rho} =0$ (where $\gamma^{\mu\nu\rho} := \gamma^{[ \mu}
\gamma^{\nu} \gamma^{\rho ]}$)
implies the Maxwell equation $\partial^{\mu} \partial_{[ \mu} \psi_{\nu 
]} =0$, and conversely
  acting on the  Maxwell equation with ${1 \over \square} \dirac$
  gives an equation equivalent to the Rarita-Schwinger
equation {\cite{keyFraSag}}.

In {\cite{keyFro}} first order field equations were given for general spinor-valued totally symmetric rank $s$ tensor gauge fields (referred to as spin-$(s+ 1/2)$ fields) in the ${[ 1,1,...,1
]}_S$ representation, which are invariant under gauge transformations 
with constrained parameters. In {\cite{keyFraSag}},
a non-local form of these equations was found which is invariant 
under gauge transformations with unconstrained parameters.
Consider the case of a spin-$5/2$ tensor field $\psi \in X^{{[1,1]}_S}$ whose first order field equation
in {\cite{keyFro}} is
%
%
\begin{equation}
\dirac \, \psi_{\mu\nu} - 2 \partial_{( \mu} {{\not}\psi}_{\nu )} \;\;
=\;\; 0
\label{eq:3.1}
\end{equation}
where ${{\not}\psi}_{\nu} := \gamma^{\mu} \psi_{\mu\nu}$.
({\ref{eq:3.1}}) is   only  gauge-invariant under
\begin{equation}
\delta
\psi_{\mu\nu} = 2 \partial_{( \mu} \xi_{\nu )}
\label{eq:3.a}
\end{equation}
if the spin-$3/2$ parameter satisfies the constraint $\gamma^{\mm} \xi_{\mm} =0$.  A non-local  fully
gauge-invariant field equation {\cite{keyFraSag}} is obtained by 
taking the linear
combination
%
%
\begin{equation}
\dirac \, \psi_{\mu\nu} - 2 \partial_{( \mu} {{\not}\psi}_{\nu )} -
{\partial_{\mu} \partial_{\nu} \over \square} \left( \dirac \,
\psi^{\rho}_{\;\; \rho} - 2 \partial^{\rho} {{\not}\psi}_{\rho} \right)
\;\; =\;\; 0
\label{eq:3.2}
\end{equation}
of ({\ref{eq:3.1}}) with its trace. This is invariant under 
({\ref{eq:3.a}})
with unconstrained parameter.
Acting on
({\ref{eq:3.2}}) with $\dirac$ one obtains the second order linearised 
Einstein
equation
%
%
\begin{equation}
\eta^{\mu\rho} \partial_{[ \mu} \psi_{\nu ][ \rho , \sigma ]} \;\;
=\;\; 0
\label{eq:3.3}
\end{equation}
which is gauge-invariant and local. Conversely, one obtains
({\ref{eq:3.2}}) from ({\ref{eq:3.3}}) by acting with ${1 \over \square} \dirac$ on the latter. The
generalisation to arbitrary spinor-valued spin-$s$ fermionic fields is
then straightforward {\cite{keyFraSag}}; one
obtains fully gauge-invariant field equations by taking non-local
linear combinations of the field equation
%
%
\begin{equation}
\dirac \, \psi_{\mu_1 ... \mu_s} - s \partial_{( \mu_1}
{{\not}\psi}_{\mu_2 ... \mu_s )} \;\; =\;\; 0
\label{eq:3.4}
\end{equation}
from {\cite{keyFro}}, where ${{\not}\psi}_{\mu_2 ... \mu_s} := \gamma^{\mu_1} \psi_{\mu_1 ... \mu_s}$.
A second order field equation is obtained by acting on
the first order gauge-invariant field equation with
$\dirac$, and this second order equation for a spinor-valued 
spin-$(s+1/2)$ fermionic field is that discussed in previous sections for a spin-$s$ 
bosonic field
but with $A$ replaced with $\psi$. The first
order field equation is regained by acting on this second order equation with ${1 \over \square} \dirac$.


\subsection{First order equations for general ${[ p_1 ,..., p_N ]}_S$
tensors}

This generalises to  general spinor-valued tensor fields.
The operations on multi-forms extend trivially to spinor-valued
multi-forms. The
local gauge-invariant action   for a fermionic field $\psi \in
X^{{[ p_1 ,..., p_N ]}_S}$ is
%
%
\begin{equation}
{\cal{S}}^{{[ p_1 ,..., p_N ]}_S} \;\; =\;\; - \left( \prod_{i=1}^{N}
\frac{1}{p_i !} \right) \int d^D x \; {\bar{\psi}}^{\; {\mu}^1_1
...{\mu}^1_{p_1} ...{\mu}^i_1 ...{\mu}^i_{p_i} ...{\mu}^N_{p_N} }
E_{{\mu}^1_1 ...{\mu}^1_{p_1} ...{\mu}^i_1 ...{\mu}^i_{p_i}
...{\mu}^N_{p_N}} ( \psi )
\label{eq:3.5}
\end{equation}
where ${\bar{\psi}}$ denotes the Dirac conjugate of $\psi$. The gauge-invariant type ${[ p_1 ,..., p_N ]}_S$
fermionic field equation tensor $E( \psi )$ involves $N$ partial
derivatives on $\psi$ for even $N$ (or $N+1$ derivatives for odd $N$)
and $E$ is identical, as an operator, to that given earlier in terms of
derivatives of $A$. In particular,  $E$ again satisfies the $N$ 
conservation
conditions ${d^\dagger}^{(i)} E \equiv 0$ identically for $i=1,...,N$.
For $N$ even, the fermionic field equation derived
from ({\ref{eq:3.5}}), for a general type ${[ p_1 ,..., p_N ]}_S$
fermion $\psi$ is given by
%
%
\begin{equation}
\sum_{I \in S_N} \tau^{(
i_1 i_2 )} ... \tau^{( i_{N-1} i_N )} F( \psi ) \;\; =\;\; 0
\label{eq:3.6}
\end{equation}
whilst for $N$ odd, the derived fermionic field equation is given by
%
%
\begin{equation}
\sum_{I \in S_N}
\tau^{( i_1 i_2 )} ... \tau^{( i_N \, N+1)} \pa F( \psi ) \;\;
=\;\; 0
\label{eq:3.7}
\end{equation}
where $F( \psi ) = d^{(1)} ... d^{(N)} \psi$ is the fermionic type ${[
p_1 +1 ,..., p_N +1 ]}_S$ tensor field strength for $\psi$.

Non-local first order field equations can be obtained from these
local higher derivative equations by acting with ${1 \over 
\square^{r}} \dirac$
for suitable $r$.
The non-local gauge-invariant action is
%
%
\begin{equation}
{\cal{S}}^{{[ p_1 ,..., p_N ]}_S}_{(r+ 1/2)} \;\; =\;\; - \left(
\prod_{i=1}^{N} \frac{1}{p_i !} \right) \int d^D x \; {\bar{\psi}}^{\;
{\mu}^1_1 ...{\mu}^1_{p_1} ...{\mu}^i_1 ...{\mu}^i_{p_i}
...{\mu}^N_{p_N} } {\dirac \over \square^{\, r+1}} E_{{\mu}^1_1
...{\mu}^1_{p_1} ...{\mu}^i_1 ...{\mu}^i_{p_i} ...{\mu}^N_{p_N}} ( \psi
)
\label{eq:3.8}
\end{equation}
where, as in ({\ref{eq:1.1}}), the power $r$ is chosen to be ${N \over
2} -1$ for $N$ even and ${(N+1) \over 2} -1$ for $N$ odd so that the
derived field equations are of first order. The  gauge-invariant  field
equations are given by
%
%
\begin{eqnarray}
{\cal{G}}^{(0)}_S &:=& \sum_{I \in S_N} \tau^{( i_1 i_2
)} ... \tau^{( i_{N-1} i_N )} {\dirac \over {\square}^{\, {N \over 2}}}
F( \psi ) \;\; =\;\; 0 \label{eq:3.9} \\ [.1in]
{\cal{G}}^{(1)}_S &:=& \sum_{I \in S_N}
\tau^{( i_1 i_2 )} ... \tau^{( i_N \, N+1)} {\dirac \over {\square}^{\,
{(N+1) \over 2}}} \pa F( \psi ) \;\; =\;\; 0 \label{eq:3.10}
\end{eqnarray}
for $N$ even and odd respectively. For spinor-valued spin-$s$ fields
(with $N=s$ and all $p_i =1$) these gauge-invariant fermionic field
equations correspond to those proposed in {\cite{keyFraSag}}.

As in the bosonic case, one can construct associated first order non-local field expressions
%
%
\begin{eqnarray}
{\cal G}^{(2n)}_S &:=& \sum_{I \in S_{N+2n}} \tau^{( i_1 i_2 )} ... \tau^{( i_{N-1+2n}
\; i_{N+2n} )} {{\dirac} \over {\square}^{\, {N \over 2}
+n}} F^{(2n)} ( \psi )  \label{eq:3.9a} \\ [.1in]
{\cal G}^{(2n+1)}_S &:=& \sum_{I \in S_{N+2n}} \tau^{( i_1 i_2 )} ... \tau^{(
i_{N+2n} \, N+1+2n)} {{\dirac} \over {\square}^{\, {(N+1) \over 2} +n}}
F^{(2n+1)} ( \psi ) \label{eq:3.10a}
\end{eqnarray}
for the case of $N$ even and odd respectively. The field equation ({\ref{eq:3.9}}) implies the vanishing of
({\ref{eq:3.9a}}) for all $n$ while ({\ref{eq:3.10}}) implies that 
({\ref{eq:3.10a}}) are zero. In general, one can also consider general gauge-invariant field equations which are the linear combinations $\sum_n a_n \, {\cal{G}}^{(2n)}_S =0$ or $\sum_n a_n \, {\cal{G}}^{(2n+1)}_S =0$ for some coeffiecients $a_n$.

The non-local action is replaced by
the local Dirac form
%
%
\begin{equation}
{\cal{S}}^{{[ p_1 ,..., p_N ]}_S}_{(r+ 1/2)} \;\; =\;\; - \left(
\prod_{i=1}^{N} \frac{1}{p_i !} \right) \int d^D x \; {\bar{\psi}}^{\;
{\mu}^1_1 ...{\mu}^1_{p_1} ...{\mu}^i_1 ...{\mu}^i_{p_i}
...{\mu}^N_{p_N} } {\dirac} \, {\psi}_{ {\mu}^1_1 ...{\mu}^1_{p_1}
...{\mu}^i_1 ...{\mu}^i_{p_i} ...{\mu}^N_{p_N} }
\label{eq:3.8a}
\end{equation}
on imposing the physical gauge conditions
%
%
\begin{equation}
{d^{\dagger}}^{(i)} \psi \;\; =\;\; 0 \quad\quad , \quad\quad \gamma^{\mu^i_1} \psi_{{\mu}^1_1 ...{\mu}^1_{p_1} ...{\mu}^i_1 ...{\mu}^i_{p_i} ... {\mu}^N_{p_N} } \;\; =\;\; 0
\label{eq:3.8b}
\end{equation}
for all $i=1,...,N$. Note that the second of these constraints implies $\tau^{(ij)} \psi  =  0$ for any $i,j =1,...,N$. This follows by multiplying the gamma-tracelessness condition by $\gamma^{\mu^j_1}$ and using $\sigma^{(ij)} \psi =0$ for $j >i$ (since the tensor part of $\psi$ is $GL(D,{\mathbb{R}})$-irreducible).

To conclude, we have
%
%
\begin{equation}
{\dirac}\, {\cal{G}}^{(m)}_S ( \psi ) \;\; =\;\; {\cal{G}}^{(m)} ( \psi
)
\label{eq:3.11}
\end{equation}
as an operator equation for any spinor-valued fermionic tensor field
$\psi$, where ${\cal{G}}^{(m)} ( \psi )$ correspond to the second order
operators defined in ({\ref{eq:1.2}}), ({\ref{eq:1.3}}),
({\ref{eq:1.4}}) and ({\ref{eq:1.5}}) but now acting on $\psi$. This
generalises the result in {\cite{keyFraSag}}.
Conversely,
%
%
\begin{equation}
   {\cal{G}}^{(m)}_S ( \psi ) \;\; =\;\;  {\dirac \over
\square} {\cal{G}}^{(m)} ( \psi
)
\label{eq:3.11a}
\end{equation}
%


\section*{Appendix A : Fixing to physical gauge for higher spins}

A free massless gauge field in $D$ dimensions can be reduced, by gauge 
fixing and using the field equations, to the dynamical degrees of 
freedom corresponding to a field in a representation of the little
group $SO(D-2) \subset SO(D-1,1)$ satisfying a free field equation.
Rather than fully fixing such a light-cone gauge, it will be sufficient 
here to consider the
analogue of the transverse traceless gauge in general relativity, 
and we shall refer to such gauges as \lq  physical gauges'.
For example, consider a free massless totally symmetric tensor gauge 
field $\phi_{\mu_1 ... \mu_s}$ of rank $s$.
The gauge symmetry can be used to impose a gauge condition such as
$\partial^{\mu_1} \phi_{\mu_1 \mu_2 ... \mu_s} = 0$ off-shell.
However, if $\phi_{\mu_1 ... \mu_s}$ satisfies its field equation,
further restricted gauge transformations are possible while preserving 
the gauge conditions and these can be used to make $\phi_{\mu_1 ... \mu_s}$ traceless, and the field 
equation then reduces to the free one $\square\, \phi_{\mu_1 ... \mu_s} =0$.
Then $\phi_{\mu_1 ... \mu_s}$ is in physical gauge if
\begin{equation}
\square\, \phi_{\mu_1 ... \mu_s} \;\; =\;\; 0 \quad\quad , \quad\quad 
\partial^{\mu_1} \phi_{\mu_1 \mu_2 ... \mu_s} \;\; =\;\; 0 \quad\quad , 
\quad\quad \eta^{\mu_1 \mu_2} \phi_{\mu_1 \mu_2 ... \mu_s} \;\; =\;\; 0
\label{eq:a1}
\end{equation}
on-shell, where $\eta^{\mu\nu}$ is the (inverse) $SO(D-1,1)$-invariant 
metric. We now discuss this gauge-fixing in more detail for the examples $s=2,3,4$.

A massless field with $s=2$ describes a linearised graviton $h_{\mu\nu}$ whose field equation is the linearised Einstein equation
\begin{equation}
{\cal{G}}_{\mu\nu} \;\; :=\;\; \square h_{\mu\nu} - 2 \partial^{\rho} 
\partial_{( \mu} h_{\nu ) \rho} + \partial_{\mu} \partial_{\nu} 
h^{\prime} \;\; =\;\; 0
\label{eq:a2}
\end{equation}
where $h^{\prime} := \eta^{\mu\nu} h_{\mu\nu}$ is the trace. ({\ref{eq:a2}}) is invariant under the gauge transformation
\begin{equation}
  \begin{array}{rcl}
  \delta h_{\mu\nu} &=& 2 \partial_{( \mu} \xi_{\nu )} \\
  \delta h^{\prime} &=& 2 \partial^{\mu} \xi_{\mu} \\
  \end{array}
\label{eq:a3}
\end{equation}
for any one-form parameter $\xi_{\mu}$. The De Donder gauge choice
\begin{equation}
T_{\mu} \;\; :=\;\; \partial^{\nu} h_{\mu\nu} - \frac{1}{2} 
\partial_{\mu} h^{\prime} \;\; =\;\; 0
\label{eq:a4}
\end{equation}
uses $D$ gauges symmetries to impose $D$ constraints, but
doesn't quite fix all the gauge symmetry, as it allows gauge transformations 
preserving the constraint $T_{\mu} =0$,
\begin{equation}
\delta T_{\mu} \;\; =\;\; \square \xi_{\mu}  \;\; =\;\; 0
\label{eq:a5}
\end{equation}
restricting the gauge transformations to those with parameters $\xi_{\mu}$ satisfying $\square \xi_{\mu}   =  0$.
For on-shell configurations satisfying ({\ref{eq:a2}}), this residual symmetry
can be used to eliminate the trace of the on-shell graviton $h^{\prime}$. The field equation ({\ref{eq:a2}}) implies
$\square h^{\prime} = 0$ (using ({\ref{eq:a4}})) and for $h^{\prime}$ satisfying this, one can solve
the equation
\begin{equation}
h^{\prime} \;\; =\;\; 2 \partial^{\mu} \zeta_{\mu}
\label{eq:a6}
\end{equation}
for some $\zeta_{\mu}$ satisfying $\square \zeta_{\mu} =0$ (see e.g. {\cite{keyWald}}), and so a gauge transformation with parameter $\xi_\mu = - \zeta _{\mu}$ can be used to set $h^{\prime} =0$ on-shell. Setting $h^{\prime} =0$ implies that 
({\ref{eq:a4}}) reduces to $\partial^{\mu} h_{\mu\nu} =0$ and ({\ref{eq:a2}}) reduces to $\square h_{\mu\nu} =0$, and the transverse traceless or physical gauge is achieved.

For $s=3$, the field equation for a massless gauge field $\phi_{\mu\nu\rho}$  is 
\begin{equation}
{\cal{G}}_{\mu\nu\rho} \;\; :=\;\; \square \phi_{\mu\nu\rho} - 3 
\partial^{\alpha} \partial_{( \mu} \phi_{\nu\rho ) \alpha} + 
\partial_{( \mu} \partial_{\nu} {\phi}^{\prime}_{\rho )} + 
\frac{1}{\square} \left( 2 \partial^{\alpha} \partial^{\beta} 
\partial_{( \mu} \partial_{\nu} \phi_{\rho ) \alpha\beta} - 
\partial_{\mu} \partial_{\nu} \partial_{\rho} \partial^{\alpha} 
\phi^{\prime}_{\alpha} \right) \;\; =\;\; 0
\label{eq:b1}
\end{equation}
where $\phi^{\prime}_{\mu} := \eta^{\nu\rho} \phi_{\mu\nu\rho}$ is the 
trace. This non-local equation is invariant under the gauge transformation
\begin{equation}
  \begin{array}{rcl}
  \delta \phi_{\mu\nu\rho} &=& 3 \partial_{( \mu} \xi_{\nu\rho )} \\
  \delta \phi^{\prime}_{\mu} &=& \partial_{\mu} \xi^{\,\prime} +2 
\partial^{\nu} \xi_{\nu\mu} \\
  \end{array}
\label{eq:b2}
\end{equation}
for any second rank symmetric tensor parameter $\xi_{\mu\nu}$. A convenient gauge choice is
\begin{equation}
T_{\mu\nu} \;\; :=\;\; \partial^{\rho} \phi_{\mu\nu\rho} - \partial_{( 
\mu} \phi^{\prime}_{\nu )} \;\; =\;\; 0
\label{eq:b3}
\end{equation}
but this still allows gauge transformations with
parameters  $\xi_{\mu\nu}$ satisfying
\begin{equation}
\delta T_{\mu\nu} \;\; =\;\; \square \xi_{\mu\nu} - \partial_{\mu} 
\partial_{\nu} \xi^{\,\prime}  \;\; =\;\; 0
\label{eq:b4}
\end{equation}
These can now be used to eliminate the trace $\phi^{\prime}_{\mu}$ provided $\phi_{\mu\nu\rho}$ satsfies the field equation ({\ref{eq:b1}}). The field equation ({\ref{eq:b1}}) and gauge condition ({\ref{eq:b3}}) imply that the trace satisfies
\begin{equation}
\square  \phi^{\prime}_{\mu} \;\; =\;\; \partial_{\mu} \partial^{\nu} \phi^{\prime}_{\nu}
\label{eq:b5}
\end{equation}
Given a second rank symmetric tensor $\zeta_{\mu\nu}$ which satisfies   
\begin{equation}
\phi^{\prime}_{\mu} \;\; =\;\; \partial_{\mu} \zeta^{\,\prime} + 2 \partial^{\nu} \zeta_{\nu\mu}
\label{eq:b6}
\end{equation}
together with
\begin{equation}
\square \zeta_{\mu\nu} - \partial_{\mu} \partial_{\nu} \zeta^{\, \prime} \;\; =\;\; 0
\label{eq:b7}
\end{equation}
one can can perform a gauge transformation with parameter $\xi_{\mu\nu} = - \zeta_{\mu\nu}$ to set $\phi^{\prime}_{\mu} =0$, so that ({\ref{eq:b3}}) reduces to $\partial^{\mu} \phi_{\mu\nu\rho} =0$ and ({\ref{eq:b1}}) reduces to $\square \phi_{\mu\nu\rho} =0$.

It remains to show that a tensor $\zeta_{\mu\nu}$ can be chosen to satisfy ({\ref{eq:b6}}), ({\ref{eq:b7}}). Define
\begin{equation}
f_\mu  \;\; :=\;\; \phi^{\prime}_{\mu} -\partial_{\mu} \zeta^{\,\prime} 
- 2 \partial^{\nu} \zeta_{\nu\mu}
\label{eq:b7x}
\end{equation}
The strategy, following {\cite{keyWald}}, is to arrange for $f_\mu$ and $\dot  
f_\mu$ to vanish on an initial value surface $t = t_0$ (where $t := x^0$ is 
the time coordinate and  $\dot g := \partial_t  g$ for any tensor $g$). Then if
$\square f_\mu =0$, $f_\mu$ will vanish everywhere and ({\ref{eq:b6}}) will hold.
Note that the trace of ({\ref{eq:b7}}) vanishes identically, so that no constraint 
is imposed on $\square \zeta^{\,\prime}$ by ({\ref{eq:b7}}). Then $\zeta^{\,\prime}$ can be  
chosen to satisfy
\begin{equation}
\square \zeta^{\,\prime}  \;\; =\;\; {1 \over 3} \partial^{\mu} \phi^{\prime}_{\mu}
\label{eq:b8}
\end{equation}
so that this and ({\ref{eq:b5}}) imply $\square f_\mu=0$ (using ({\ref{eq:b7}})), so that $f_\mu$ is 
harmonic when $\phi^{\prime}_{\mu}$ is on-shell.
Then ({\ref{eq:b7}}) becomes the following constraint on ${\hat{\zeta}}_{\mu\nu}$, 
the trace-free part of $\zeta_{\mu\nu}$,
\begin{equation}
\square {\hat{\zeta}}_{\mu\nu} - \partial_{\mu} \partial_{\nu} \zeta^{\, 
\prime} \;\; =\;\;
-{1 \over D} \eta_{\mu\nu} \partial^{\rho} \phi^{\prime}_{\rho}
\label{eq:b9}
\end{equation}
On the initial value surface $t= t_0$ we choose $\zeta_{\mu\nu}$, $\dot  
\zeta_{\mu\nu}$ to satisfy
\begin{equation}
\phi^{\prime}_{\mu} \;\; =\;\; \partial_{\mu} \zeta^{\,\prime} + 2 \partial^{\nu} \zeta_{\nu\mu} 
\label{eq:b10}
\end{equation}
and
\begin{equation}
\dot \phi^{\prime}_{\mu} \;\; =\;\; 3 \partial_{\mu} \dot \zeta^{\,\prime} + 2 \nabla ^{i} \dot 
\zeta_{i\mu} -2 \nabla^2 \zeta_{0\mu} 
\label{eq:b11}
\end{equation}
where $x^i$ with $i=1,...,D-1$ are the spatial coordinates and 
$\nabla^2 := \nabla^i \nabla_i$.
These ensure that $f_\mu =0$ and $\dot f_\mu=0$ on $t = t_0$.
Then $\zeta_{\mu\nu}$ is chosen to satisfy ({\ref{eq:b7}}) and ({\ref{eq:b9}}); given the 
initial values of $\zeta_{\mu\nu}$, $\dot \zeta_{\mu\nu}$ at $t= t_0$, this 
determines $\zeta_{\mu\nu}$ uniquely.
Then as $\square f_\mu =0$, it follows that $f_\mu =0$ everywhere, and 
as a result ({\ref{eq:b6}}) and ({\ref{eq:b7}}) are indeed satisfied, as required.

For $s=4$, a massless field $\phi_{\mu\nu\rho\sigma}$ can satisfy the field equation
\begin{equation}
  \begin{array}{rcl}
  {\cal{G}}_{\mu\nu\rho\sigma} &:=& \square \phi_{\mu\nu\rho\sigma} - 4 
\partial^{\alpha} \partial_{( \mu} \phi_{\nu\rho\sigma ) \alpha} + 2 
\partial_{( \mu} \partial_{\nu} {\phi}^{\prime}_{\rho\sigma )} \\
  &&+ \frac{1}{\square} \left( 4 \partial^{\alpha} \partial^{\beta} 
\partial_{( \mu} \partial_{\nu} \phi_{\rho\sigma ) \alpha\beta} - 4 
\partial_{( \mu} \partial_{\nu} \partial_{\rho} \partial^{\alpha} 
\phi^{\prime}_{\sigma ) \alpha} + \partial_{\mu} \partial_{\nu} 
\partial_{\rho} \partial_{\sigma} \phi^{\prime\prime}\right) \;\; =\;\; 
0 \\
  \end{array}
\label{eq:c1}
\end{equation}
where $\phi^{\prime}_{\mu\nu} := \eta^{\rho\sigma} \phi_{\mu\nu\rho\sigma}$ and $\phi^{\prime\prime} := \eta^{\mu\nu} 
\phi^{\prime}_{\mu\nu}$ are the single and double traces. The non-locality in ({\ref{eq:c1}}) is again necessary so that it is invariant under the gauge transformation
\begin{equation}
  \begin{array}{rcl}
  \delta \phi_{\mu\nu\rho\sigma} &=& 4 \partial_{( \mu} 
\xi_{\nu\rho\sigma )} \\
  \delta \phi^{\prime}_{\mu\nu} &=& 2 \partial_{( \mu} 
\xi^{\,\prime}_{\nu )} +2 \partial^{\rho} \xi_{\mu\nu\rho} \\
  \delta \phi^{\prime\prime} &=& 4 \partial^{\mu} \xi^{\,\prime}_{\mu}
  \end{array}
\label{eq:c2}
\end{equation}
with unconstrained third rank totally symmetric tensor parameter $\xi_{\mu\nu\rho}$. The gauge constraint
\begin{equation}
T_{\mu\nu\rho} \;\; :=\;\; \partial^{\sigma} \phi_{\mu\nu\rho\sigma} - 
\frac{3}{2} \partial_{( \mu} \phi^{\prime}_{\nu\rho  )} \;\; =\;\; 0
\label{eq:c3}
\end{equation}
restricts the gauge transformations to those with parameters $\xi_{\mu\nu\rho}$ satisfying
\begin{equation}
\delta T_{\mu\nu\rho} \;\; =\;\; \square \xi_{\mu\nu\rho} - 3 
\partial_{( \mu} \partial_{\nu} \xi^{\,\prime}_{\rho )}  \;\; =\;\; 0
\label{eq:c4}
\end{equation}
These can be used to set the single trace $\phi^{\prime}_{\mu\nu} $ to zero when
$\phi_{\mu\nu\rho\sigma}$ satisfies its field equations. The field equation ({\ref{eq:c1}}) and gauge condition ({\ref{eq:c3}}) imply that the trace satisfies
\begin{equation}
\square \phi^{\prime}_{\mu\nu} \;\; =\;\; 2 \partial^{\rho} \partial_{( \mu} \phi^{\prime}_{\nu )\rho}
\label{eq:c5}
\end{equation}
and that
\begin{equation}
\partial_{\mu} \phi^{\prime\prime} \;\; =\;\; 0 \qquad , \qquad 
\partial^{\mu} \partial^{\nu} \phi^{\prime}_{\mu\nu} \;\; =\;\; 0
\label{eq:c5a}
\end{equation}

A third rank totally symmetric tensor $\zeta_{\mu\nu\rho}$ satisfying
\begin{equation}
\phi^{\prime}_{\mu\nu} \;\; =\;\; 2 \partial_{( \mu} \zeta^{\,\prime}_{\nu )} + 2 \partial^{\rho} \zeta_{\mu\nu\rho}
\label{eq:c6}
\end{equation}
and
\begin{equation}
\square \zeta_{\mu\nu\rho} - 3 \partial_{( \mu} \partial_{\nu} \zeta^{\, \prime}_{\rho )}  \;\; =\;\; 0
\label{eq:c7}
\end{equation}
can then be used as a parameter of a gauge transformation with $\xi_{\mu\nu\rho} = - \zeta_{\mu\nu\rho}$ that sets $\phi^{\prime}_{\mu\nu} =0$, so that ({\ref{eq:c3}}) reduces to $\partial^{\mu} \phi_{\mu\nu\rho\sigma} =0$ and ({\ref{eq:c1}}) reduces to $\square \phi_{\mu\nu\rho\sigma} =0$. 

As before, such a tensor $\zeta_{\mu\nu\rho}$ can be found by first specifying $\zeta_{\mu\nu\rho}$, $\dot \zeta_{\mu\nu\rho}$ on an initial value surface $t= t_0$ and then using a wave equation for $\zeta_{\mu\nu\rho}$ to fix the tensor 
everywhere. Note that the trace of ({\ref{eq:c7}}) does not restrict $\square \zeta^{\,\prime}_{\mu}$, but does imply
\begin{equation}
\partial_{\mu} \partial^{\nu} \zeta^{\,\prime}_{\nu}  \;\; =\;\; 0
\label{eq:c8}
\end{equation}
Equation ({\ref{eq:c6}}) implies 
\begin{equation}
  \phi^{\prime\prime} \;\; =\;\; 4
 \partial^{\nu} \zeta^{\,\prime}_{\nu} 
  \label{eq:c7a}
\end{equation}
and both sides of this equation are constant, as a result of ({\ref{eq:c5a}}), ({\ref{eq:c8}}).
Defining
\begin{equation}
f_{\mu\nu} \;\; :=\;\;  \phi^{\prime}_{\mu\nu} -2 \partial_{( \mu} 
\zeta^{\,\prime}_{\nu )} - 2 \partial^{\rho} \zeta_{\mu\nu\rho}
\label{eq:c9}
\end{equation}
then if $\zeta^{\,\prime}_{\mu}$ is chosen to satisfy
\begin{equation}
\square \zeta^{\,\prime}_{\mu} \;\; =\;\; {1 \over 3} \partial^{\nu} \phi^{\prime }_{\mu\nu}
\label{eq:c10}
\end{equation}
it follows that $\square f_{\mu\nu} =0$ on-shell. As above, $\zeta_{\mu\nu\rho}$, $\dot \zeta_{\mu\nu\rho}$ can be chosen 
at $t= t_0$ so that $f_{\mu\nu} =0$ and $\dot f_{\mu\nu} =0$ at $t= t_0$, and these 
together with ({\ref{eq:c7}}), ({\ref{eq:c10}}) determine $\zeta_{\mu\nu\rho}$ everywhere. It then follows that
$f _{\mu\nu} =0$ everywhere and so ({\ref{eq:c6}}) and ({\ref{eq:c7}}) are satisfied, as required.


\section*{Appendix B : The generalised Einstein tensor for type $[ p_1 ,..., p_N ]$ gauge fields}

As has been seen, the quadratic action for a type $[ p_1 ,..., p_N ]$  gauge field  is naturally written in terms of a generalised Einstein tensor $E$ which is a gauge-invariant type $[ p_1 ,..., p_N ]$  tensor that is conserved (i.e.   ${d^\dagger}^{(i)} E \equiv 0$ for all $i=1,...,N$). It is straightforward to construct $E$ in simple examples such as those discussed in the introduction. In this appendix we use the multi-form structure to write an explicit form for $E$ in some simple cases and the leading terms in the general case. Based on these results we propose an expression for the form of $E$ in the general case.
 
Before discussing the construction for general $N$, it will prove useful to present the details for the $N=2$ case. This class of bi-form gauge theories is illustrative of the general structure. In this case, we can use the simplified notation of {\cite{keythesis}}, {\cite{keydeMHul}}, dropping the superscript $(1)$ and replacing the the superscript $(2)$ with a tilde, so that $d:= d^{(1)}$, ${\tilde{d}} := d^{(2)}$, and omitting the superscript $(12)$, so that, e.g. $\tau := \tau^{(12)}$. We begin by noting the following identities for bi-form operators acting on a general element $T \, \in \, X^{p,q}$
%
%
\begin{eqnarray}
d \tau^n + {(-1)}^{n+1} \tau^n d &=& n\, {\tilde{d^\dagger}} \tau^{n-1} \nonumber \\
d^\dagger \tau + \tau d^\dagger &=& 0 \nonumber \\
d \eta + \eta d &=& 0 \nonumber \\
d^{\dagger} \eta^n + {(-1)}^{n+1} \eta^n d^{\dagger} &=& n\, {\tilde{d}} \eta^{n-1} \label{eq:x1} \\
\tau \eta - \eta \tau &=& (D-p-q) 1 \nonumber \\
\sigma \tau &=& \tau \sigma \nonumber \\
\sigma \eta &=& \eta \sigma \nonumber
\end{eqnarray}
with similar relations holding for the operators with tildes. 

The general form for $E$ can be written as
%
%
\begin{equation}
E \;\; = \;\; \sum_{n=0}^{q} k_n \; \eta^n \tau^{n+1} F
\label{eq:x2a}
\end{equation}
for some coefficients $k_n$, and these coefficients are fixed by requiring that $E$ be conserved.
For a gauge field $A \, \in \, X^{[p,q]}$ (with $p \geq q$) with field strength $F = d{\tilde{d}} A \, \in \, X^{[p+1,q+1]}$, the identities ({\ref{eq:x1}}) imply that $\sigma \eta^m \tau^n F =0$ for any powers $m$ and $n$ (since $\sigma F =0$). 
This implies that each term in the sum ({\ref{eq:x2a}}) is annihilated by $\sigma$ and so as a result each term in the sum is $GL(D,{\mathbb{R}})$-irreducible, in the $[p,q]$ representation.
The identity 
%
%
\begin{equation}
d^{\dagger} ( \eta^n \tau^{n} {\cal{G}} ) \;\; = \;\; \eta^n \tau^n d^{\dagger} {\cal{G}} + n(n+1)\, \eta^{n-1} \tau^{n-1} d^{\dagger} {\cal{G}}
\label{eq:x3}
\end{equation}
allows the coefficients in ({\ref{eq:x2a}}) to be determined order by order in the expansion in the powers of $\eta$ by requiring conservation of $E$ at each order. Requiring that $E$ satisfies $d^\dagger E \equiv 0$ and ${\tilde{d^\dagger}} E \equiv 0$ identically fixes the coefficients, giving the result 
%
%
\begin{equation}
E \;\; = \;\; \sum_{n=0}^{q} {{(-1)}^n \over (n+1){(n!)}^2} \; \eta^n \tau^{n+1} F
\label{eq:x2}
\end{equation}
and $E$ is in the $[p,q]$ representation, as required. The generalised Einstein equation $E =0$ implies ${\cal{G}} := \tau F =0$ for $D> p+q$. For example, for linearised gravity, $p=q=1$ and the usual Einstein equation $E_{\mu\nu} =0$ implies that the Ricci tensor vanishes in dimensions $D>2$, but in the critical dimension $D=2$, $E_{\mu\nu}=0$ is an identity implying no restriction on the Ricci tensor.

We now turn to the general case of multi-form gauge fields $A \, \in \, X^{[ p_1 ,..., p_N ]}$.
The following identities for operators acting on a general element $T \, \in \, X^{p_1 ,..., p_N}$ will be useful 
%
%
\begin{eqnarray}
d^{(j)} \tau^{(ij)} + \tau^{(ij)} d^{(j)} &=& {d^\dagger}^{(i)} \nonumber \\
{d^\dagger}^{(j)} \tau^{(ij)} + \tau^{(ij)} {d^\dagger}^{(j)} &=& 0 \nonumber \\
d^{(j)} \eta^{(ij)} + \eta^{(ij)} d^{(j)} &=& 0 \nonumber \\
{d^{\dagger}}^{(j)} \eta^{(ij)} + \eta^{(ij)} {d^{\dagger}}^{(j)} &=& d^{(i)} \nonumber \\
\tau^{(ij)} \tau^{(kj)} + \tau^{(kj)} \tau^{(ij)} &=& 0 \nonumber \\
\eta^{(ij)} \eta^{(kj)} + \eta^{(kj)} \eta^{(ij)} &=& 0 \nonumber \\
\sigma^{(ij)} \tau^{(ij)} - \tau^{(ij)} \sigma^{(ij)} &=& 0 \label{eq:x4} \\
\sigma^{(ij)} \tau^{(jk)} + \tau^{(jk)} \sigma^{(ij)} &=& 0 \nonumber \\
\sigma^{(ji)} \tau^{(jk)} + \tau^{(jk)} \sigma^{(ji)} &=& - \tau^{(ik)} \nonumber \\
\sigma^{(ij)} \eta^{(ij)} - \eta^{(ij)} \sigma^{(ij)} &=& 0 \nonumber \\
\sigma^{(ij)} \eta^{(jk)} + \eta^{(jk)} \sigma^{(ij)} &=& - \eta^{(ik)} \nonumber \\
\sigma^{(ji)} \eta^{(jk)} + \eta^{(jk)} \sigma^{(ji)} &=& 0 \nonumber
\end{eqnarray}
Operators with distinct labels commute. Repeated labels are not to be summed. 

The field equation for even  $N$ was given by ({\ref{eq:4.25}}) (or (45)) involving multiple traces of $F$, while that for odd  $N$ was
formally very similar, given by  ({\ref{eq:4.24}}) (or (46))
but with $F$ replaced by the tensor $\partial F$, corresponding to a tableau with an even number  ($N+1$)  of columns. 
Below we will discuss the case of even $N$; similar formulae can be used for the odd $N$ case provided 
$F$ is  replaced by  $\partial F$.

{\textbf{Theorem :}} For a gauge field $A \, \in \, X^{[ p_1 ,..., p_N ]}$ (with $p_i \geq p_{i+1}$) with field strength $F = d^{(1)} ... d^{(N)} A \, \in \, X^{[ p_1 +1 ,..., p_N +1]}$, the identities ({\ref{eq:x4}}) imply that 
%
%
\begin{equation}
\sigma^{(ij)} \; \sum_{I \in S_N} \tau^{( i_1 i_2 )} ... \tau^{( i_{N-1} i_N )} F \;\; =\;\; 0
\label{eq:x5}
\end{equation}
since $\sigma^{(ij)} F =0$ for any $j>i$. The sum is over all values of $I = ( i_1 ,..., i_N )$ in $( 1,...,N)$ with no $i_k$ equal. There are 
%
%
\begin{equation}
f_N \;\; :=\;\; {N! \over 2^{N/2} \, (N/2)!}
\label{eq:x5n}
\end{equation}
such inequivalent terms.
\footnote{Two terms are said to be inequivalent if their indices cannot be rearranged such that they are proportional to each other.}

{\textbf{Proof :}} Begin by partitioning the sum in ({\ref{eq:x5}}) into two seperate sums for any given $\sigma^{(ij)}$. The first sum ${\cal{G}}_1$ contains $f_{N-2}$ inequivalent terms whose elements each have one $\tau^{(ij)}$ in the $(N/2)$-fold trace of $F$. The second sum ${\cal{G}}_2$ contains the remaining $f_N - f_{N-2}$ inequivalent terms whose elements each have one $\tau^{(i i_r )}$ and one $\tau^{(j i_s )}$ where $i_r$ and $i_s$ are different labels for different terms but never equal $i$ nor $j$. Since each of the $N/2$ traces have different labels then they commute and can be arbitrarily permuted. We therefore choose each $\tau^{(j i_s )}$ to be leftmost in ${\cal{G}}_2$. From ({\ref{eq:x4}}) it is clear that $\sigma^{(ij)}$ commutes with all the traces in ${\cal{G}}_1$. Moreover, $\sigma^{(ij)}$ commutes with all traces in ${\cal{G}}_2$ except the two $\tau^{(j i_s )}$ and $\tau^{(i i_r )}$ traces. ({\ref{eq:x4}}) shows that $\sigma^{(ij)}$ anticommutes with the first of these traces $\tau^{(j i_s )}$ then anticommutes with the second $\tau^{(i i_r )}$ but also produces another term with $\tau^{( j i_r )}$ replacing each $\tau^{(i i_r )}$. More precisely this means  
%
%
\begin{eqnarray}
& \sigma^{(ij)} & \sum_{I \in S_N} \tau^{( i_1 i_2 )} ... \tau^{( i_{N-1} i_N )} F - \sum_{I \in S_N} \tau^{( i_1 i_2 )} ... \tau^{( i_{N-1} i_{N} )} \sigma^{(ij)}\, F \nonumber \\
&& \label{eq:x5a} \\  
&=& \sum \tau^{(j i_s )} \tau^{(j i_r )} \tau^{( i_1 i_2 )} ... \tau^{( i_{N-1} i_N )} F
\nonumber
\end{eqnarray}
where the sum on the right hand side is still over all labels $i_k \neq i,j$ with $k=1,...,s,...,r,...,N$. However, the first two traces in this sum have a common index $j$ and therefore anticommute (using ({\ref{eq:x4}})). Since one sums over all inequivalent labels then for each $i_s =m$ and $i_r =n$, there will be a corresponding pair $i_s =n$ and $i_r =m$ and so this sum is identically zero. Using that $F$ is irreducible under $GL(D,{\mathbb{R}})$ then completes the proof.$\;\;\; {\bf{\square}}$

The $GL(D,{\mathbb{R}})$-irreducible term
%
%
\begin{equation}
{\cal{G}} \;\; := \;\; \sum_{I \in S_N} \tau^{( i_1 i_2 )} ... \tau^{( i_{N-1} i_N )} F
\label{eq:x6}
\end{equation}
is the leading term in $E \, \in \, X^{[ p_1 ,..., p_N ]}$. The first shifted trace term is given by  
%
%
\begin{equation}
\eta \cdot \tau {\cal{G}} \;\; :=\;\; {\cal{Y}}_{[ p_1 ,..., p_N ]} \circ \left( {1 \over 2} \sum_{i,j =1}^{N} \eta^{(ij)} \tau^{(ij)} \right) {\cal{G}}
\label{eq:x7}
\end{equation}
with relative coefficient $k_1 = -1/N$ chosen to ensure conservation of $E \, =\, {\cal{G}} - (1/N) \eta \cdot \tau {\cal{G}}+\dots $ to first order. This coefficient is computed using the relation
%
%
\begin{equation}
{d^{\dagger}}^{(i)} ( \eta \cdot \tau {\cal{G}} ) \;\; = \;\; \eta \cdot \tau\, {d^{\dagger}}^{(i)} {\cal{G}} + N\, {d^{\dagger}}^{(i)} {\cal{G}} 
\label{eq:x7a}
\end{equation}
A natural guess for the next shifted trace term is that it should be proportional to
%
%
\begin{equation}
\eta^2 \cdot \tau^2 {\cal{G}} \;\; :=\;\; {\cal{Y}}_{[ p_1 ,..., p_N ]} \circ \left( {1 \over 4} \sum_{i,j,k,l =1}^{N} \left( \eta^{(ij)} \eta^{(kl)} + 2 \eta^{(ik)} \eta^{(lj)} \right) \tau^{(ij)} \tau^{(kl)} \right) {\cal{G}}
\label{eq:x8}
\end{equation}
Similarly, a natural proposal for the general $n$-trace correction term is that it be proportional to
%
%
\begin{equation}
\eta^n \cdot \tau^n {\cal{G}} \;\; :=\;\; {\cal{Y}}_{[ p_1 ,..., p_N ]} \circ \left( {1 \over 2^n} \sum_{i_1 ,..., i_{2n} =1}^{N} \left( \sum_\pi \eta^{( \pi ( i_1 ) \pi ( i_2 ))} ... \eta^{( \pi ( i_{2n-1} ) \pi ( i_{2n} ))} \right) \tau^{( i_1 i_2 )} ... \tau^{( i_{2n-1} i_{2n} )} \right) {\cal{G}}
\label{eq:x9}
\end{equation}
The bracketed sum in ({\ref{eq:x9}}) is over all permutations of labels $( i_1 ,..., i_{2n} )$. Those permutations which occur more than once should be counted with multiplicity (hence the factor of 2 in the second term in ({\ref{eq:x8}})).  
In general, an explicit Young projection onto the $[ p_1 ,..., p_N ]$ representation is required, although it can be the case, as in the bi-form examples above, that it is not needed as the term is already irreducible without projection.

This leads to the conjecture that $E$ is given by
%
%
\begin{equation}
E \;\; = \;\; \sum_{n} k_n \; \eta^n \cdot \tau^{n} {\cal{G}}
\label{eq:x10}
\end{equation}
with coefficients $k_n$ chosen so that ${d^\dagger}^{(i)} E \equiv 0$. It is clear, however, that the structure of the higher trace correction terms is complicated and this makes the determination of coefficients $k_n$ difficult for $n>1$. 

The ordering of algebraic operators in ({\ref{eq:x10}}) (with all $\eta$'s to the left) is chosen because ${d^{\dagger}}^{(i)} ( \eta^n \cdot \tau^n {\cal{G}} )$ can be expressed as terms with only $n$ and $n-1$ traces of ${d^\dagger}^{(i)} {\cal{G}}$ (rather than a sum of terms with all possible traces). This follows from the identities ({\ref{eq:x4}}). If our conjectured form for the shifted traces is correct, then it is to be expected that there
should be a general relation
%
%
\begin{equation}
{d^{\dagger}}^{(i)} ( \eta^n \cdot \tau^n {\cal{G}} ) \;\; = \;\; \eta^n \cdot \tau^n \, {d^{\dagger}}^{(i)} {\cal{G}} + {\textsf{P}}_N (n) \; \eta^{n-1} \cdot \tau^{n-1} \, {d^{\dagger}}^{(i)} {\cal{G}}
\label{eq:x11}
\end{equation}
for some polynomial ${\textsf{P}}_N (n)$ in $N$ and $n$. This structure was found above for ${\textsf{P}}_N (1) = N$ ({\ref{eq:x7a}}) and ${\textsf{P}}_2 (n) = n(n+1)$ ({\ref{eq:x3}}) and will be checked in further examples below.  If ({\ref{eq:x11}}) is true then conservation of $E$ implies    
%
%
\begin{equation}
k_n \;\; = \;\; {(-1)}^n \, {\left( \prod_{r=1}^n {\textsf{P}}_N (r) \right)}^{-1}
\label{eq:x12}
\end{equation}
after choosing $k_0 =1$.

A check of the conjecture ({\ref{eq:x11}}) is given by considering the example of totally symmetric rank $s$ gauge fields. We 
again take $s$ to be even; the odd spin case is similar but with
$F$ replaced by $\partial F$. As noted in the introduction, the gauge-invariant field strength $F$ for a spin-$s$ (type $[1,...,1]$ tensor) gauge field is a type $[2,...,2]$ tensor. The corresponding gauge-invariant field equation is defined using ({\ref{eq:x6}}) and is given by the vanishing of the spin-$s$ tensor ${\cal{G}}$ with components    
%
%
\begin{equation}
{\cal{G}}_{\mu_1 ... \mu_s} \;\; := \;\; f_s \, \eta^{( \nu_1 \nu_2} ... \eta^{\nu_{s-1} \nu_s )} F_{\mu_1 \nu_1 ... \mu_s \nu_s}
\label{eq:x13}
\end{equation}
where $f_s$ is the numerical factor defined in ({\ref{eq:x5n}}) so that each inequivalent term in ({\ref{eq:x13}}) contributes with weight 1. The shifted trace terms are defined using ({\ref{eq:x9}}) and are also spin-$s$ tensors whose components are  
%
%
\begin{equation}
{({\eta^n \cdot \tau^n {\cal{G}}})}_{\mu_1 ... \mu_s} \;\; := \;\; g_{s,n} \, \eta_{( \mu_1 \mu_2} ... \eta_{\mu_{2n-1} \mu_{2n}} \, {\cal{G}}^{\{ n \}}_{\mu_{2n+1} ... \mu_s )}
\label{eq:x14}
\end{equation}
where ${\cal{G}}^{\{ n \}} := \tau^n {\cal{G}}$ is the $n$th trace of ${\cal{G}}$ for $0 \leq n \leq s/2$. There is just a single inequivalent trace $\tau$ on totally symmetric tensors. The numerical factor 
%
%
\begin{equation}
g_{s,n} \;\; := \;\; {s! \over n! \, 2^n \,  (s-2n)!}
\label{eq:x15}
\end{equation}
ensures that each inequivalent term in ({\ref{eq:x14}}) contributes with weight 1. 
The shifted trace terms ({\ref{eq:x14}}) for spin-$s$ gauge fields are irreducible, without the need for an explicit Young projection -- in contrast to the general case discussed above. This is due to the definition of the sum over permutations in ({\ref{eq:x9}}) which is equivalent to a total symmetrisation of indices when acting on a totally symmetric tensor.

Using the relation
%
%
\begin{equation}
\partial^\mu {\cal{G}}^{\{ n-1 \}}_{\mu \mu_{2n} ... \mu_s} \;\; = \;\; {(s+1-2n) \over (s+2-2n)} \, \partial_{( \mu_{2n}} \, {\cal{G}}^{\{ n \}}_{\mu_{2n+1} ... \mu_s )}
\label{eq:x16}
\end{equation}
then implies
%
%
\begin{equation}
{d^{\dagger}} ( \eta^n \cdot \tau^n {\cal{G}} ) \;\; = \;\; \eta^n \cdot \tau^n \, {d^{\dagger}} {\cal{G}} + {\textsf{P}}_s (n) \; \eta^{n-1} \cdot \tau^{n-1} \, {d^{\dagger}} {\cal{G}}
\label{eq:x17}
\end{equation}
where ${\textsf{P}}_s (0) =0$ and ${\textsf{P}}_s (n) = s+2-2n$ for $0<n \leq s/2$. These numbers agree with those found above for ${\textsf{P}}_N (n)$ for those values of $(N,n)$ for which they are both defined. That is, one finds ${\textsf{P}}_s (1) =s$ from either ({\ref{eq:x17}}) or from ({\ref{eq:x7a}}) and ({\ref{eq:x3}}). One can then solve for the spin-$s$ Einsten tensor with even $s$ in the manner described above, so that
%
%
\begin{equation}
E \;\; = \;\; {\cal{G}} + \sum_{n=1}^{s/2} { {(-1)}^n \over \left( \prod_{r=1}^{n} (s+2-2r) \right)} \; \eta^n \cdot \tau^{n} {\cal{G}}
\label{eq:x18}
\end{equation}
which satisfies the single inequivalent conservation condition $d^\dagger E \equiv 0$ identically.


\vspace*{.4in}
{\textbf{{\large{Acknowledgements}}}}

We would like to thank M. Vasiliev for helpful comments and useful discussions.


\end{document}